\documentclass[preprintnumbers, floatfix, preprintnumbers, letterpaper, superscriptaddress,nofootinbib]{revtex4}
\usepackage{graphicx}
\usepackage{microtype}
\usepackage{amsmath}
\usepackage{amssymb}
\usepackage{subfigure}
\usepackage{hyperref}
\usepackage{url}
\usepackage{xcolor}
\usepackage{color}
\usepackage{mathrsfs}
\usepackage{calrsfs}
\usepackage{amsfonts}
\usepackage{lipsum}
\usepackage{eufrak}
\usepackage{tabularx}
\usepackage{eucal}
\usepackage{latexsym}
\usepackage{ragged2e}
\usepackage{epsfig}
\usepackage{textcomp}

\usepackage{caption}
\DeclareCaptionJustification{justified}{\leftskip=0pt \rightskip=0pt \parfillskip=0pt plus 1fil}
\definecolor{vividviolet}{rgb}{0.62, 0.0, 1.0}
\definecolor{amaranth}{rgb}{0.9, 0.17, 0.31}
\definecolor{palatinateblue}{rgb}{0.15, 0.23, 0.89}
\definecolor{brightpink}{rgb}{1.0, 0.0, 0.5}
\definecolor{cornflowerblue}{rgb}{0.39, 0.58, 0.93}
\definecolor{deepcarminepink}{rgb}{0.94, 0.19, 0.22}
\definecolor{radicalred}{rgb}{1.0, 0.21, 0.37}

\hypersetup{ linktoc=all,
	colorlinks, linkcolor={palatinateblue},
	citecolor={brightpink}, urlcolor={amaranth}
}

\graphicspath{{Images/}}

\renewcommand{\d}[1]{\ensuremath{\operatorname{d}\!{#1}}}

\def\sideremark#1{\ifvmode\leavevmode\fi\vadjust{\vbox to0pt{\vss
			\hbox to 0pt{\hskip\hsize\hskip1em
				\vbox{\hsize1.3cm\tiny\raggedright\pretolerance10000
					\noindent #1\hfill}\hss}\vbox to8pt{\vfil}\vss}}}%
\def\beq{\begin{equation}}
\def\eeq{\end{equation}}

\begin{document}

\title{On the Chandrasekhar Limit in Generalized Uncertainty Principles}

\author{Daniele \surname{Gregoris}}
\email{danielegregoris@libero.it}
\affiliation{Department of Physics, School of Science, Jiangsu University of Science and Technology, Zhenjiang 212003, China}

\author{Yen Chin \surname{Ong}}
\email{ycong@yzu.edu.cn}
\affiliation{Center for Gravitation and Cosmology, College of Physical Science and Technology, Yangzhou University, \\180 Siwangting Road, Yangzhou City, Jiangsu Province  225002, China}
\affiliation{Shanghai Frontier Science Center for Gravitational Wave Detection, School of Aeronautics and Astronautics, Shanghai Jiao Tong University, Shanghai 200240, China}

\begin{abstract}
The Chandrasekhar limit for white dwarfs has been confirmed by many astrophysical observations. However, how to obtain it theoretically in models which rely on other-than-Heisenberg's uncertainty principles, which are predicted by some quantum gravity theories, is not a trivial task. 
In this manuscript, we will derive the Chandrasekhar mass assuming an uncertainty relation proposed in the framework of Doubly Special Relativity, exhibiting both a minimal length and a maximum momentum.   We will show how to  re-obtain an asymptotically vanishing radius for the star in the limiting subcase of the Heisenberg uncertainty principle:
 this is the same behavior as in the original Chandrasekhar's derivation, but \emph{not} when the more popular Generalized Uncertainty Principle is assumed. 
We will argue that this is related to the difference in the \emph{classical} limit of these two different generalized uncertainty principles,   enlightening that the modified uncertainty parameter considered here behaves like a finite temperature.
The role of a maximum momentum in guaranteeing the stability of the configuration,  and in providing a finite pressure for the degenerate electrons' gas,  is also analyzed. 
\end{abstract} 

\maketitle

\section{Introduction}

The three fundamental constants $\hbar$, $c$ and $G$ represent the realm of quantum mechanics (via Heisenberg's uncertainty  principle), special relativity (via the dispersion relation between energy, mass and momentum), and gravity (via the Newton's force law).  In the modern formulation according to which the latter two theories are merged, both $c$ and $G$ enter Einstein's equations of general relativity, which up to date have passed various observational tests, with remarkable precisions.  On the other hand, there is still no universal consensus regarding if -- and how -- $c$ and $G$ should play some role in modifying Heisenberg's uncertainty principle. According to the so-called ``Generalized Uncertainty Principle'' (GUP) the Heisenberg principle should be modified as
\beq
\label{dquantum}
\Delta x \Delta p \sim \frac{\hbar}{2}[1+ \beta^2 (\Delta p)^2]\,,
\eeq
where $\beta=1/(M_p c)$ with $M_p=\sqrt{\frac{\hbar c}{G}}$ being the Planck mass; switching off gravitational effects by considering $G \to 0$ the textbook uncertainty relation is reproduced. One advantage of this proposal is that a minimal length naturally arises as postulated by various quantum gravity theories/models \cite{gupi1,gupi2,gupi3,gupi4,gupi5,paff}. Interestingly, this modification of quantum mechanics affects the Hawking evaporation of black holes in such a way that stable remnants may exist \cite{gupi6}, and even constitute viable candidates for dark matter \cite{gupi7}. Since laboratory settings are far from probing this energy regime, astrophysical phenomena  constitute the only available arena with recent claims that the observed baryon asymmetry may be due to GUP-induced effects \cite{baryo}, while another signature would be a weaker stochastic gravitational wave signal sourcing from the phase transition in the early universe \cite{gw}.  

Furthermore, one possible remedy for the Greisen-Zatsepin-Kuzmin (GZK) anomaly in the production of cosmic rays  through particles collisions with the cosmic microwave background radiation might be by requiring also a maximum value for the momentum \cite{amelino0}. A specific uncertainty relation which accounts for  the existence of both a minimal length and  a maximum momentum has been constructed in \cite{pedram1,pedram2} and slightly modified  recently in \cite{highergup} to an expression which is not restricted to even functions of the momentum; in this latter paper  the toy-models of the one-dimensional box and harmonic oscillator problems have been studied. The existence of  maximum/minimal values for the momentum/length may potentially cure also the IR/UV divergences that plague the field theory computation of the value of the cosmological constant \cite{weincc}. It is clear that meaningful extensions of the Heisenberg principle should not only tame open tensions but also being consistent with all the previously well established physics. For example, when describing the matter content of white dwarfs and neutron stars adopting the GUP,  (re-)obtaining the maximum value for the stellar mass has  required some not-so-trivial thinking \cite{ong1,ong2,annals,rsos}. This is a crucial test that newly formulated theories need to pass because a maximum value for such masses has been theoretically predicted by various authors like Chandrasekhar \cite{original1,original2}, Landau \cite{landau}, Stoner \cite{stoner}, Tolman-Oppenheimer-Volkoff \cite{tov1,tov2} and confirmed observationally \cite{annu1,annu2}. Therefore, an assessment of the uncertainty relation proposed in \cite{highergup} with respect to these astrophysical requirements is in order. We refer the reader also to \cite{gen2021a,gen2021b,gen2021c,gen2021d,gen2021e} for some further investigations of the properties of stellar configurations in gravitational equilibrium in other-than-Heisenberg models. 

The major concern here pertains to a peculiar phenomenon that was soon noticed when applying GUP to white dwarf physics -- the Chandrasekhar limit seems not to exist anymore under GUP correction; see \cite{ong1, annals}. Among the possible resolutions to this puzzle is to take into account general relativistic effects properly. The reason is simple: if quantum gravitational effect like GUP is already important, then surely general relativity needs to be considered. In fact it was already suggested in \cite{ong1} that the GUP-corrected Tolman-Oppenheimer-Volkoff (TOV) equation would be required to fully
model the stars; though in that work it was instead suggested as a possible solution that perhaps the GUP parameter is negative.
Subsequently, in \cite[Sect.3.1.2]{rsos} it was explained how to obtain the maximum value for the mass of a white dwarf in the high-momentum limit by implementing the equation of state of a degenerate electron gas, which is sensitive to the GUP corrections, into the Tolman-Oppenheimer-Volkoff equations; the procedure worked as follows: the mass-radius relation was obtained, the maximum radius of the configuration was estimated and then used in that equation to get the maximum value of the mass. Although this result has resolved the tension between astrophysical observations and  some GUP models, it may not fully reflect Chandrasekhar's way of thinking who emphasized that his mass limit actually arises when the size of the white dwarf is \emph{vanishing}\footnote{A body with a vanishing size and  a finite mass has infinite  density and is thus unphysical. However, the point is that this is a \emph{limiting} configuration that is not attainable. In the regime of ultra-relativistic electrons, should the star get compressed,  the matter pressure and the force of gravity  increase at the same rate safeguarding the equilibrium condition and forcing the star to become arbitrarily small \cite{garf}.  However, the Chandrasekhar limit constitutes an asymptotic state beyond which the matter pressure would be so strong to make the star collapse into a black hole \cite{garf}; this is consistent with astrophysical observations which point out that the mass of white dwarfs are {\it below} and not {\it at} the Chandrasekhar limit. See more comments in the Discussion section \ref{ss6}. } \cite{original2}. In this paper, we will show that following  the same method of \cite{rsos}, but adopting the uncertainty principle called GUP* formulated in \cite{highergup}, it is possible to re-obtain the Chandrasekhar mass consistently with his point of view.  In fact, we will show  that in the limiting subcase in which corrections beyond the Heisenberg uncertainty principle  are switched off, the radius becomes zero. We will claim as well that another advantage of considering a formalism exhibiting a maximum momentum is that the pressure of the degenerate electrons' gas in not anymore diverging in the high-momentum limit, consequently implying that the gravitational pressure balancing it is always finite as well.
 
Another objective of our work is to show that just like GUP, the newly proposed GUP* also allows white dwarf solutions once general relativistic effect is properly considered, and so cannot be ruled out at this stage as a possible viable candidate of modified uncertainty principle. There are, however, noticeable differences in the two modifications, as already mentioned.
The Chandrasekhar mass is a consequence or manifestation of microscopic (quantum) effects in the macroscopic world;  the fact that this limit behaves qualitatively differently when GUP and GUP* are adopted may be related to their  classical limit $\hbar \to 0$ being \emph{different}. Indeed, if one sets $\beta=1/(M_p c)$, the term $\hbar \beta^2  \to \text{const.} \neq 0$ and for (\ref{dquantum}) we get {$\Delta x \sim G \Delta p/(2c^3)$}, while the classical limit of the uncertainty relation proposed in (\cite{highergup}) is $\Delta x \Delta p \sim 0$, as is also the case for the original Heisenberg one. Although this may represent a tiny  discrepancy from the numerical perspective ($\hbar$ being tiny anyway), it constitutes a clear conceptual difference. We will return to this issue in the Discussion section \ref{ss6}. 

Our distinctive assumption of a maximum momentum for the electrons constituting the white dwarfs can also be interpreted from the following perspective. As the electrons get more and more packed during the compression of the star, their interaction force $F$, which generates the pressure counter-balancing gravity, increases. Here we are assuming that this interaction force should not diverge. Recalling that $F= \d p /\d t$, and writing $F=p_{\rm max}/t_p = \left( \frac{\hbar c}{G} \right)^{1/2} \cdot c \cdot \left( \frac{\hbar G}{c^5} \right)^{-1/2} =F_{\rm max}$, where the latter is, up to a $O(1)$ proportionality factor, the maximum force conjectured by Gibbons \cite{gibbons1,gibbons2} and Schiller \cite{shiller1,shiller2}. The maximum force conjecture implies that only a maximum amount of power  can be radiated \cite{dyson}, and indeed white dwarfs explode into supernovae which are standard candles \cite{annu1,annu2}.  

Our paper is organized as follows: in Sect. \ref{ss2} we review some basic properties of the uncertainty principle we are adopting and quantify its effects on the number density, energy density and pressure of a degenerate electron gas. In Sects. \ref{ss3} and \ref{ss4} we inspect the condition for having a gravitational equilibrium within \emph{Newtonian} gravity and assuming the electrons to be non-relativistic and relativistic particles, respectively,  and the consequences of the existence or lack of a maximum momentum are examined. The purpose here is to make clear comparisons with the previous literature in the context of GUP. In Sect. \ref{ss5} we move to the general relativistic formulation using the GUP corrected Tolman-Oppenheimer-Volkoff equations; we will obtain the Chandrasekhar limit and comment on the crucial difference with the previously-considered GUP formalism -- notably, in our case the size of the star is asymptotically vanishing; we also establish the stability of the configurations.  In Sect. \ref{ss6a} we will show that our main conclusion on the existence of the Chandrasekhar limit still holds also in a class of modified theories of gravity in which the Vainshtein screening mechanism is broken, and we will argue that  the specific deviation from the Heisenberg principle assumed in this paper behaves like a temperature effect.  We conclude in Sect. \ref{ss6} summarizing our work and putting it in the perspective of possible future investigations.

\section{Introducing the GUP* in the context of white dwarf }
\label{ss2}

In \cite{highergup} the following higher order generalized uncertainty principle (GUP*) has been proposed:
\beq
\label{newgup}
\Delta x \Delta p \sim \frac{\hbar}{2}\left[-\beta \Delta p +\frac{1}{1- \beta \Delta p}  \right]\,,
\eeq
which, among other properties,  comes with a maximal momentum and a minimal length\footnote{Another model with these properties was recently explored in \cite{gen2021e}.}.  It is straightforward to check that in the classical limit for which $\hbar \to 0$, using $\beta=1/(M_p c)$, we obtain $\Delta x \Delta p \sim 0$, just like the usual Heisenberg's uncertainty principle.
This uncertainty relation has been derived from the position-momentum commutation rule
\beq
\label{commutation}
[\hat x,\, \hat p]= \frac{i \hbar}{1-\beta p}\,,
\eeq
which, should one choose a positive $\beta$, comes with the maximum momentum $p_{max} =1/\beta$ for the particle it describes, which is consistent with the requirements of the theory of Doubly Special Relativity \cite{amelino1,amelino2,amelino3,amelino4}.
In \cite{jeans}, corrections to the Jeans mass predicted by such a new form of uncertainty principle have been computed claiming that it affects the estimate of the potential energy in a way that may account for the gravitational collapse and  star formation in the Bok globules \cite{Bok}. It is therefore interesting to examine other astrophysical consequences of GUP* to see if it is indeed viable.

Solving (\ref{newgup}) we can obtain two roots for the momentum:
\beq
\label{solp}
\Delta p \sim \frac{1}{2 \beta} \left[ 1 \pm \sqrt{\frac{2 \Delta x -3 \beta \hbar}{ 2 \Delta x +\beta \hbar }} \right]\,.
\eeq
Thus, if we choose the solution with the + square root we necessarily need to require $\beta$ to be positive and $\Delta x > 3 \beta \hbar/2$ (the latter for having a positive argument in the square root), while if we pick the solution with the $-$ sign we will have the following possibilities: if we choose $\beta>0$ we would need  the argument of the square root to be smaller than one which is automatically guaranteed as long as $\Delta x > 3 \beta \hbar/2$, while for a negative $\beta$ such argument should be larger than 1 which would deliver just the restriction $\Delta x > -\beta \hbar/2$ since the denominator should be positive. We summarize in Table \ref{TableI} the restrictions for the applicability of the two roots in (\ref{solp}).

\begin{table}
	\begin{center}
		\begin{tabular}{|c|c|}
			\hline
			\multicolumn{2}{|c|} {\text {  }}	\\[-0.7em]
			\multicolumn{2}{|c|} {\text { Sign + in (\ref{solp}) }}   \\
			[2 pt]
			\hline
		&	\\ [-0.7em]
			$\beta>0$    & $\Delta x >3\beta \hbar/2$    \\
			[2 pt]
			$\beta<0$ &  Not applicable  \\
			[2 pt]
			\hline
				\multicolumn{2}{|c|} {\text {  }}	\\[-0.7em]
				\multicolumn{2}{|c|} {\text { Sign $-$ in (\ref{solp}) }}   \\
				[2 pt]
				\hline
				&	\\ [-0.7em]
				$\beta>0$    & $\Delta x >3\beta \hbar/2$    \\
				[2 pt]
				$\beta<0$ &  $\Delta x >-\beta \hbar/2$  \\
				[2 pt]
				\hline
		\end{tabular}
		\caption{The Table summarizes the ranges of applicability of the two roots in (\ref{solp}) depending on the sign of the free parameter $\beta$; the existence of a minimal $\Delta x$ is pointed out when appropriate. }
		\label{TableI}
	\end{center}
\end{table}

At the level of first approximation we can describe a white dwarf as having a volume $V \sim R^3$ with $R$ being its radius, and having a number density $n:=N/V=M/(m_e V)\sim  \Delta x^{-3}$, where $M$ is its mass and $m_e$ the mass of an electron. The gravitational energy of the white dwarf is
\beq
\label{binding1}
E_g \sim -\frac{GM^2}{R}.
\eeq

\subsection{Exploring the GUP* Corrections to the Equation of State of a Degenerate Electrons Gas}
\label{sectgas}

\begin{figure}
	\begin{center}
		\includegraphics[scale=0.8]{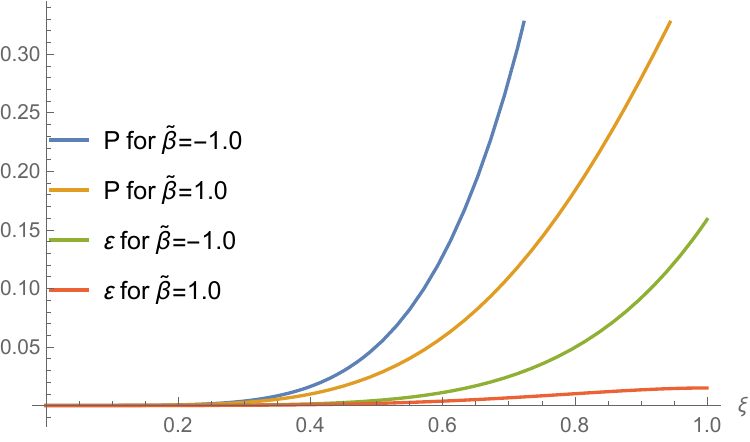}
	\end{center}
	\caption{ The figure confirms that the internal kinetic energy (\ref{kinetic}) and pressure (\ref{pressure}) computed in the modified momentum space obeying GUP* are positive; from the top to bottom, we have $P$ with $\tilde \beta=-1.0$, $P$ with $\tilde \beta=1.0$, $\varepsilon$ with $\tilde \beta=-1.0$, $\varepsilon$ with $\tilde \beta=1.0$, respectively. For graphical convenience the pressure is shown in units of $\alpha=20$, while the internal kinetic energy in units of $\alpha=1$.   }
	\label{fig1}
\end{figure}

The material of which white dwarfs are made can be approximated as a degenerate gas of electrons (fermions) at temperature $T\sim 0$. The uncertainty principle affects the properties of such a gas, like the density of particles, its pressure and total energy, which may have further consequences on the condition of gravitational equilibrium governing the existence of the star \cite{rsos,gen2021a,gen2021b,gen2021c}.  These quantities can be computed from the grand canonical	partition function as done in \cite{annals} for GUP, just by modifying the size of the phase space volume accordingly.  
Taking into account the correction to the phase space volume \cite[Eq.(9)]{phasespace} implied by the commutation relation (\ref{commutation}), the number density is found to be
\begin{eqnarray}
\label{ngeneral}
n &=& \frac{1}{\pi^2 \hbar^3}\int_0^{p_F} p^2 (1-\beta p) dp = \frac{p_F^3 (4-3\beta p_F)}{12 \pi^2 \hbar^3} = \frac{ \xi^3 \alpha (4 -3 \tilde \beta \xi )}{12 m_e c^2 },
\end{eqnarray}
where $p_F$ is the Fermi momentum, and where in the last step we have defined\footnote{While $\xi$ and $\beta$ are dimensionless, $[\alpha]=M/(L T^2)$.}  $\xi:=p_F/(m_e c)$,  $\tilde \beta :=m_e c\beta$ and  $\alpha:=m_e^4 c^5/(\hbar^3 \pi^2)$. We note that although the commutation relation (\ref{commutation}) is different than the one arising in the context of the more well-known generalized uncertainty principle GUP, for the scenario consistent with doubly special relativity requirements ($\beta>0$), they both have the effect of rendering the momentum phase space volume smaller \cite{liouv}.   Moreover, the pressure of the degenerate fermions gas would be given by 
\beq
P = \frac{1}{\pi^2 \hbar^3}\int_0^{p_F} p^2 (1-\beta p) (E_F - E_p) dp\,,
\eeq
where
\beq
\label{defenergy}
E_F= \sqrt{(c p_F)^2 + (m_e c^2)^2}\,, \qquad E_p= \sqrt{(c p)^2 + (m_e c^2)^2}\,.
\eeq
An explicit computation delivers the result
\begin{eqnarray}
\label{pressure}
P &= &   \frac{2 c \left[ \frac{15 (m_e c)^4}{16} \ln \frac{\sqrt{(m_e c)^2 +p_F^2}+p_F}{m_e c}+\left( \frac{5 p_F^3}{8} -\frac{3 \beta p_F^4}{8}+\frac{p_F (8 \beta p_F -15) (m_e c)^2}{16} -\beta (m_e c)^4 \right) \sqrt{(m_e c)^2 +p_F^2} +\beta (m_e c)^5  \right]}{15 \pi^2 \hbar^3} \nonumber\\
\label{pressureb}
&=& \frac{\alpha }{20}\left[\frac{5}{2} \ln(\sqrt{\xi^2 +1} +\xi) -\left( \tilde \beta \xi^4 -\frac{4 \tilde \beta \xi^2}{3}  -\frac{5 \xi^3}{3}  +\frac{8 \tilde \beta}{3}  +\frac{5 \xi}{2}  \right) \sqrt{\xi^2 +1}+\frac{8 \tilde \beta}{3}      \right] \,.
\end{eqnarray}
Moreover, the internal kinetic  energy of the degenerate electron gas is given by
\begin{eqnarray}
\label{kinetic}
\varepsilon &=& \frac{1}{\pi^2 \hbar^3}\int_0^{p_F} p^2 (1-\beta p) (E_p -m_e c^2) dp \\
&=& \frac{c \left[\chi \sqrt{p_F^2 +(m_e c)^2} -15 (m_e c)^4 \ln \frac{\sqrt{(m_e c)^2 +p_F^2} +p_F}{m_e c} -16 m_e c \left(\beta (m_e c)^4 -\frac{15 \beta p_F^4}{8}+ \frac{5 p_F^3}{2} \right) \right]}{120 \pi^2 \hbar^3} \\
\label{kineticb}
&=& \left[\frac{\tilde \beta \xi^4}{4} -\frac{\xi^3}{3}-\frac{2 \tilde \beta}{15}-\frac{\ln(\sqrt{\xi^2 +1}+\xi)}{8}-\left(\frac{\tilde \beta}{5} \left(\xi^4 +\frac{\xi^2 -2}{3} \right) -\frac{\xi}{4} \left( \xi^2 +\frac{1}{2} \right)\right) \sqrt{1+\xi^2}   \right] \alpha      \,; \\
\chi &=& 8 \beta [2 (m_e c)^4 -(m_e c p_F)^2 - 3 p_F^4] +15 p_F [(m_e c)^2 +2 p_F^2]\,.
\end{eqnarray}
The energy density is given by the sum of the kinetic contribution and of the rest mass density $\rho_0= m_u \mu_e n(\xi)$ as:
\beq
\label{entot}
\tilde \varepsilon (\xi) = \rho_0(\xi) c^2 + \varepsilon(\xi)\,,
\eeq
where $m_u= 1.66 \times 10^{-24}$ g is the atomic mass unit (also known as a ``dalton'') and $\mu_e=A/Z$ with $A$ and $Z$ being the mass and atomic numbers respectively. In what follows, we will denote $j:=m_u \mu_e/m_e$.
Finally, the  relativistic adiabatic index $\gamma$ for the degenerate electrons gas taking into account the corrections given by the GUP* is given by
\begin{eqnarray}
\gamma &=& \frac{ \tilde \varepsilon +P}{P}\frac{d P}{d \tilde  \varepsilon} \\
&=&\frac{5 \xi^5 (3 \tilde \beta \xi -4)^2}{\left[ 30 \ln (\sqrt{\xi^2 +1} +\xi)  -2\left( 6 \tilde \beta \xi^4-8 \tilde \beta \xi^2 -10 \xi^3 +16 \tilde \beta +15 \xi \right) \sqrt{\xi^2 +1} +32 \tilde \beta \right] (1-\tilde \beta \xi)\sqrt{\xi^2 +1}}\,, 
\end{eqnarray}
where we have computed the derivative by applying the chain rule via the quantity $\xi$, and where it should be remarked that the contribution of the rest mass density drops out from the final result. We can note that the corrections to the number density, pressure and kinetic energy are linear in the GUP* parameter $\tilde \beta$, while for the relativistic adiabatic index at the leading order we have
\begin{eqnarray}
\label{napprox}                               
\gamma & \approx & \frac{8 \xi^5\left[15\left(1-\frac{\tilde \beta \xi}{2} \right)  \ln(\sqrt{\xi^2 +1} +\xi)  +\left(\tilde\beta \xi^4 -\frac{\tilde \beta \xi^2}{2}+10 \xi^3 +16 \tilde \beta -15 \xi \right) \sqrt{\xi^2 +1} -16 \tilde \beta\right]}{15 [\xi (2 \xi^2 -3) \sqrt{\xi^2 +1} +3 \ln(\sqrt{\xi^2 +1}+\xi) ]^2 \sqrt{\xi^2 +1}}  +O(\tilde \beta^2)\,. 
\end{eqnarray}
We should also appreciate some drastic differences in the behavior at high Fermi momentum $\xi \to \infty$ of the degenerate electron gas in the framework of the GUP* as compared to that of the GUP explored in (\cite{rsos,gen2021a,gen2021b,gen2021c}): {\it (i)} the number density no longer approaches a constant but is rather diverging as it would be predicted by the usual Heisenberg's uncertainty principle; {\it (ii)} the same  would be for the internal kinetic energy; {\it (iii)} the pressure is diverging faster compared to the usual GUP approach;  {\it (iv)} for the relativistic adiabatic index we obtain $\gamma \to 5/4 \approx 1.25$ for $\xi \to \infty$, which is in better agreement with the result of an ideal gas $\gamma_{\rm ideal}=4/3 \approx 1.333$ than the result of $\gamma \approx \frac{\pi \tilde \beta^3}{16}$ found in the context of GUP. More quantitatively, we should note that the highest power term in $\xi$ may  actually be diminished by some factor in $\tilde \beta$, and thus the asymptotic expression to consider are
\begin{eqnarray}
\label{largepressure}
P_{\xi \to \infty}  \approx   \frac{ \alpha \xi^4}{4} \left( \frac{1}{3} -\frac{\tilde \beta \xi}{5} \right)    \,, \qquad
\varepsilon_{\xi \to \infty}  \approx   \frac{ \alpha \xi^4 (5 -4 \tilde \beta \xi)}{20}    \,, \qquad \tilde \varepsilon_{\xi \to \infty}  \approx   \frac{ \alpha \xi^4  }{4}\left[1-\left( j +\frac{4 \xi}{5} \right) \tilde \beta  \right]\,,
\end{eqnarray}
which for the case consistent with doubly special relativity requirements ($\beta>0$) are well-defined only if $\tilde \beta \xi =\beta p_F <5/4$, since the internal kinetic energy and pressure should be positive. We see that this condition is automatically satisfied because $p<1/\beta$ is required by the commutation relation (\ref{commutation}). It should be noted also that for both $p_F \to 1/\beta$ and $p_F \to 0$, Eq.(\ref{ngeneral}) becomes $n \propto p_F^3$, which is the relation we will use in Sects. \ref{ss3}-\ref{ss4}, while in Sect. \ref{ss5} we will employ the parametric GUP*-corrected equation of state $P=P(\tilde \varepsilon)$.
We should also note that in the small $\xi$ limit the subleading term is further suppressed by the smallness of the GUP* parameter:
\begin{eqnarray}
P_{\xi \to 0}  \approx   \frac{ \alpha \xi^5}{3} \left( \frac{1}{5} -\frac{\tilde \beta \xi}{8} \right)    \,, \qquad
\varepsilon_{\xi \to 0}  \approx   \frac{ \alpha \xi^5}{2} \left( \frac{1}{5} -\frac{\tilde \beta \xi}{6} \right) \,, \qquad
\tilde \varepsilon_{\xi \to 0}  \approx c^2 \rho_{ 0\xi \to 0}  \approx  \frac{ j \xi^3  \alpha}{3}     \,.
\end{eqnarray}
Thus, the electron rest mass is dominating the energy budget in the low momentum limit with respect to the contribution of the internal kinetic energy, but it may bring forth a correction in the high momentum approximation.
In Fig. (\ref{fig1}) we confirm that both the pressure and the internal kinetic energy, which are sensitive to the GUP*, are positive, and therefore can be applied in the modeling of the interior matter of a white dwarf. In particular, our new equation of state fulfills the weak energy condition ($\tilde \varepsilon \geqslant 0$, $\tilde \varepsilon +P \geqslant 0$) which guarantees that in the TOV framework for hydrostatic equilibrium (see Eqs.(\ref{tovm1})-(\ref{tovm2})) the pressure is monotonically decreasing from the center of the star to its boundary while the mass is increasing.

\section{Non-relativistic regime}
\label{ss3}

In the non-relativistic regime, the total kinetic energy of a white dwarf is
\beq
E_k = \frac{N \Delta p^2}{2 m_e}\sim \frac{N}{8 \beta^2 m_e} \left[ 1 \pm \sqrt{\frac{2 \Delta x -3 \beta \hbar}{ 2 \Delta x +\beta \hbar }} \right]^2 \sim \frac{M}{8 (\beta m_e)^2} \left[ 1 \pm \sqrt{\frac{2 R \, m_e^{1/3} -3\beta \hbar \, M^{1/3}}{2R \, m_e^{1/3} +\beta \hbar \, M^{1/3}}} \right]^2 \,,
\eeq
where in the second step we have used (\ref{solp}) which implements the deviations from the standard Heisenberg's uncertainty principle. For having a configuration in  gravitational equilibrium, the kinetic energy of  electrons should balance the gravitational energy of the star leading to the condition
\begin{eqnarray}
\label{mrimplicit1}
&& R \left[ 1 \pm \sqrt{\frac{2 R \, m_e^{1/3} -3\beta \hbar \, M^{1/3}}{2R \, m_e^{1/3} +\beta \hbar \, M^{1/3}}} \right]^2 -8G M (\beta m_e)^2 \approx 0   \qquad \Leftrightarrow \\
&& \sqrt{R}\left | 1 \pm \sqrt{\frac{2 R \, m_e^{1/3} -3\beta \hbar \, M^{1/3}}{2R \, m_e^{1/3} +\beta \hbar \, M^{1/3}}}  \right| \approx 2\sqrt{2GM }\beta m_e \qquad \Leftrightarrow  \nonumber\\
&&  4GM (\beta m_e)^2 (2R m_e^{1/3} + \beta \hbar M^{1/3}) -R (2R m_e^{1/3} -\beta \hbar M^{1/3}) \approx \pm R  \sqrt{(2 R \, m_e^{1/3} -3\beta \hbar \, M^{1/3})(2R m_e^{1/3} +\beta \hbar M^{1/3})},  \nonumber
\end{eqnarray}
which constitutes an implicit mass-radius relation. This is in fact a third degree equation in $R$:
\begin{eqnarray}
\label{cubic1}
\sum_{i=1}^3 a_i R^i & \approx & 0 \,, ~\text{where}\\
a_3 &=& 8 G M m_e^{8/3}  \,, \nonumber \\
a_2 & =& - (4 GM \beta m_e^{7/3})^2 - (\hbar M^{1/3})^2 \,, \nonumber \\
a_1 &=& -2G \beta^2 \hbar M^{5/3} (8 G \beta M^{2/3} m_e^{13/3} +\hbar m_e^2) \,, \nonumber \\
a_0 &=&  - (2 G\beta^2 M^{4/3} \hbar m_e^2)^2 \,. \nonumber
\end{eqnarray}
This mass-radius relation holds as long as (\ref{mrimplicit1}) is well-defined, that is, the argument of the square root is positive and the sign on the two sides match each other. The well-posedness of the square root provides the lower bound $R > \frac{3}{2} \beta \hbar ({M}/{m_e})^{1/3}$ for a positive $\beta$ and $R > -\frac{1}{2} \beta \hbar ({M}/{m_e})^{1/3}$ otherwise (these conditions are equivalent to those in Table \ref{TableI} for a well-posed higher order uncertainty principle GUP*, but written in terms of $R$ and $M$ rather than $\Delta x$). 

Next, if we choose a plus sign in front of the square root we need to have $2 m_e^{1/3}R^2 -\beta (8GM\beta m_e^{7/3} + \hbar M^{1/3})R -4GM^{4/3}\beta^3 m_e^2 \hbar <0$, which delivers
\begin{eqnarray}
&& \frac{8GM \beta m_e^{7/3} + \hbar M^{1/3}- \tilde s}{4 m_e^{1/3}}\beta<R<\frac{8GM \beta m_e^{7/3} + \hbar M^{1/3}+ \tilde s}{4 m_e^{1/3}}\beta \qquad  {\rm for} \, \beta>0 \,, \\
&& \frac{8GM \beta m_e^{7/3} + \hbar M^{1/3}+ \tilde s}{4 m_e^{1/3}}\beta<R<\frac{8GM \beta m_e^{7/3} + \hbar M^{1/3}- \tilde s}{4 m_e^{1/3}}\beta \qquad  {\rm for} \, \beta<0 \,; \\
\label{eqs}
&& \tilde s = \sqrt{(8GM\beta m_e^{7/3} + \hbar M^{1/3})^2 +32G\hbar \beta (m_e^7 M^4)^{1/3} }\,,
\end{eqnarray}
where the condition for $\beta<0$ does not apply because of the discussion reported in Table \ref{TableI}. For the same reason we understand that the upper bound in the condition for the radius of the white dwarf for $\beta>0$ is well-defined (e.g. positive), and therefore we have a well-posed mass-radius relation in that range. On the other hand, should we choose the minus sign in front of the square root, we would need to consider
\begin{eqnarray}
&& R<\frac{8GM \beta m_e^{7/3} + \hbar M^{1/3}- \tilde s}{4 m_e^{1/3}}\beta \qquad {\rm or} \qquad R>\frac{8GM \beta m_e^{7/3} + \hbar M^{1/3}+ \tilde s}{4 m_e^{1/3}}\beta \qquad  {\rm for} \, \beta>0 \,, \\
&& R<\frac{8GM \beta m_e^{7/3} + \hbar M^{1/3}+ \tilde s}{4 m_e^{1/3}}\beta\qquad {\rm or} \qquad R>\frac{8GM \beta m_e^{7/3} + \hbar M^{1/3}- \tilde s}{4 m_e^{1/3}}\beta \qquad  {\rm for} \, \beta<0 \,.
\end{eqnarray}
We should note that for a positive $\beta$ the radius $R$ would be negative in the former range, and similarly for a negative $\beta$. Moreover, about the case $\beta>0$ we note that the condition 
\beq
\frac{8GM \beta m_e^{7/3} + \hbar M^{1/3}+ \tilde s}{4 m_e^{1/3}}\beta >\frac{3}{2}\beta \hbar \left(\frac{M}{m_e}\right)^{1/3}\,,
\eeq
would give
\beq
\tilde s > 5 \hbar M^{1/3} - 8  GM \beta m_e^{7/3} \,,
\eeq
which is trivially fulfilled. On the other hand, for a negative value of $\beta$ we can see that $-\frac{1}{2} \beta \hbar \left(\frac{M}{m_e} \right)^{1/3}>\frac{8GM \beta m_e^{7/3} + \hbar M^{1/3}- \tilde s}{4 m_e^{1/3}}\beta$ as it would imply\footnote{Here we used the fact that multiplying both sides by a negative quantity ($\beta$) reverses the direction of the inequality.} $3\hbar M^{1/3} +8GM\beta m_e^{7/3}>\tilde s$, which then would give $-8M^{2/3}\hbar^2<0$. More importantly,  unlike  the case of a positive $\beta$ the existence of the square root is not trivially guaranteed and it can be used for constraining the mass of the white dwarf. Setting $y=M^{2/3}$, this condition becomes
\beq
(8G m^{7/3} \beta y)^2 + 48 G m^{7/3}\beta h y +h^2>0\,,
\eeq
which delivers
\beq
\label{eqy}
0<y<\frac{(3-2\sqrt{2})\hbar}{8 m^{7/3} |\beta| G} \qquad {\rm or } \qquad y>\frac{(3+2\sqrt{2})\hbar}{8 m^{7/3} |\beta| G} \,.
\eeq
We will further discuss in what follows the importance of the former range.
The applicability conditions for the mass-radius relation (\ref{cubic1}) are summarized in Table \ref{TableII}. 

\begin{table}
	\begin{center}
		\begin{tabular}{|c|c|}
			\hline
		 \multicolumn{2}{|c|} {\text {  }}	\\[-0.7em]
			 \multicolumn{2}{|c|} {\text { Condition on the Radius }}   \\
			[2 pt]
			\hline
				&	\\ [-0.7em]
			 &  Sign + in (\ref{mrimplicit1})    \\
			 [2 pt]
			\hline
				&	\\ [-0.7em]
		$\beta>0$    & $\frac{3}{2} \beta \hbar ({M}/{m_e})^{1/3}<R<\frac{8GM \beta m_e^{7/3} + \hbar M^{1/3}+ \tilde s}{4 m_e^{1/3}}\beta$    \\
			[2 pt]
			\hline
			\multicolumn{2}{|c|} {\text {  }}	\\[-0.7em]
			\multicolumn{2}{|c|} {\text { Sign $-$ in (\ref{mrimplicit1}) }}   \\
			[2 pt]
			\hline
			&	\\ [-0.7em]
			$\beta>0$    &    $R>\frac{8GM \beta m_e^{7/3} + \hbar M^{1/3}+ \tilde s}{4 m_e^{1/3}}\beta$  \\
		[2 pt]
		\hline
		&	\\ [-0.7em]
		$\beta<0$    &    $R > -\frac{1}{2} \beta \hbar ({M}/{m_e})^{1/3}$   \\
		[2 pt]
		\hline
		\end{tabular}
		\caption{The Table summarizes the ranges of applicability of the mass-radius relation (\ref{cubic1}) for a white dwarf in gravitational equilibrium in the non-relativistic regime assuming the validity of the GUP* (\ref{newgup}). The definition of $\tilde s$ is given in (\ref{eqs}). }
		\label{TableII}
	\end{center}
\end{table}

Considering a small value for $\beta$, the mass-radius relation (\ref{cubic1}) can be approximated as
\beq
8GM m_e^{8/3} R^2 -[(4GM\beta m_e^{7/3})^2 +(\hbar M^{1/3})^2]R -2GM^{5/3} (\beta \hbar m_e)^2 \approx 0\,,
\eeq
which is an even function in $\beta$ (i.e. its algebraic sign does not matter).
Two mathematical solutions can be found
\beq
R_{1,2} \approx \frac{(4GM \beta m_e^{7/3})^2 +(M^{1/3} \hbar)^2 \pm \sqrt{[(4GM \beta m_e^{7/3})^2 +(M^{1/3} \hbar)^2]^2 + (8 G M^{4/3} m_e^{7/3} \beta \hbar)^2}}{16 GM m_e^{8/3}}  \,,
\eeq
of which it can be easily seen without any ambiguity that the physically relevant one is $R_1$, the other being negative; we also note that conditions on the existence of the square root cannot be used for estimating the mass of the white dwarf because it is trivially real.  The acceptable solution therefore belongs to the branch in which we choose a minus sign in front of the square root entering (\ref{mrimplicit1}) as seen from the discussion in Table \ref{TableII}, and it is consistent with the well-known estimate \cite{ong1}
\beq
R :=R_E \sim \frac{\hbar^2}{G M^{1/3} m_e^{8/3}}
\eeq
in the limit $\beta \to 0$. More quantitatively, the corrections to the uncertainty principle encoded in (\ref{newgup}) implies a larger radius for the white dwarf than the one obtained when adopting  Heisenberg's uncertainty relation:
\beq
\label{issue}
R \approx R_E +GM (2m_e \beta)^2\,,
\eeq
which is an even function in the parameter $\beta$ whose sign therefore does not matter.
Similarly to what was obtained in \cite{ong1,annals} when the model was based on the GUP, we get again $\lim_{M \to \infty } R(M) = \infty$. Thus,  this result would suggest to assume a negative $\beta$ and consider the upper bound for the mass of the white dwarf found in  (\ref{eqy}). This conclusion is, however, only valid for non-general relativistic discussions.

\section{Relativistic regime}
\label{ss4}

In the relativistic regime the kinetic energy of the electrons is dominating over their rest masses and the condition for having a gravitational equilibrium becomes:
\beq
\label{eqrelativ}
|E_g| \sim \frac{GM^2 }{R}  \sim E_k \sim  Nc \Delta p  \sim \frac{c M}{2 \beta m_e} \left[ 1 \pm \sqrt{\frac{2 \Delta x -3 \beta \hbar}{ 2 \Delta x +\beta \hbar }} \right]  \sim \frac{c M}{2 \beta m_e} \left[ 1 \pm \sqrt{\frac{2 R m_e^{1/3} -3 \beta \hbar M^{1/3}}{ 2R m_e^{1/3} +\beta \hbar M^{1/3}}} \right]\,,
\eeq
or equivalently
\beq
\label{eq37}
2 \beta G M m_e -cR \sim \pm cR \sqrt{\frac{2 R m_e^{1/3} -3 \beta \hbar M^{1/3}}{ 2R m_e^{1/3} +\beta \hbar M^{1/3}}}\,.
\eeq
We note that the existence of the square root provides the same restrictions already discussed for the non-relativistic regime; moreover in the range $R<2 \beta G M m_e/c$ (which, consistent with Table \ref{TableI},  holds only for positive $\beta$) we should consider the plus sign in front of the square root, while for $R>2 \beta G M m_e/c$ the minus sign provides the applicable equation. For avoiding our configuration to describe a black hole we should also require that\footnote{Following \cite{rgupc}, we consider the radius of a Schwarzschild black hole to be related to its mass as $r_s \sim 2M$ also when (\ref{newgup}) holds. We would like to mention as well that following the same reference, we can write $\Delta x \sim r_s \sim 2M$ and $\Delta p \sim M$, and from the uncertainty relation (\ref{newgup}) we would get a third degree equation for $M$ whose real solution approximated at small $\beta$  delivers $M \sim 1/\beta$ meaning that a consequence of this specific modification to  Heisenberg's uncertainty relation is the prediction of the existence of black holes with Planck mass, because from \cite{highergup} we should consider $\beta=1/M_p c$. } $2 GM/c^2 <R<2 \beta GM m_e /c$, which yields $\beta > 1/(m_e c)  $, which is inconsistent with the requirement of $\beta \lesssim  1/(M_p c)$ with $M_p$ the Planck mass \cite{highergup}; we will in any case explore further the case of the plus sign for pointing out further shortcomings in its application.

In the relativistic case, considering $\beta \to 0$, we can approximate (\ref{eq37}) as
\begin{eqnarray}
\label{cond1}
&& -\frac{c M}{\beta m_e} +\left[GM^2 +\frac{c\hbar}{2}\left(\frac{M}{m_e} \right)^{4/3}  \right]\frac{1}{R}+O(\beta^2)\sim 0 \qquad {\rm for \,\, plus} \\
\label{cond2}
&& \left[GM^2 -\frac{c\hbar}{2}\left(\frac{M}{m_e} \right)^{4/3}  \right]\frac{1}{R} +O(\beta^2)\sim 0 \qquad {\rm for \,\, minus}\,.
\end{eqnarray}
Therefore, in the former case the mass-radius relationship does not contain the Chandrasekhar mass in the leading order of $\beta$, unlike  the latter which instead is consistent with it. 
The  result in (\ref{cond1})  is at odd with astrophysical observations which  have been pointing out a sharp radius-independent value for the mass of white dwarfs in gravitational equilibrium known as the Chandrasekhar limit; in fact, an almost constant absolute magnitude measured in different supernovae explosions can be regarded as an evidence that the amount of gravitational energy released is almost the same in all those events \cite{annu1}. The lack of the Chandrasekhar limit in our formalism is a direct consequence of having used an other-than-Heisenberg's uncertainty principle. In fact, if one uses the usual relationship $\Delta x \Delta p \sim \hbar$  in (\ref{eqrelativ}), then the following result would be found \cite{original1,original2}
\beq
\label{cmass}
\frac{G M^2}{R}\sim \frac{M^{4/3}\hbar c}{m_e^{4/3}R} \qquad \Rightarrow \qquad M_{\rm Ch}\sim \frac{1}{m_e^2}\left( \frac{\hbar c}{G} \right)^{3/2}\,.
\eeq
Similarly, also the assumption of the GUP
\beq
\label{gupeq}
\Delta x \Delta p \sim \hbar +\frac{\alpha L_p^2}{\hbar}(\Delta p)^2\,,
\eeq
where $L_p$ denotes the Planck length, in this method delivers a result inconsistent  with astrophysical observations \cite{ong1}.  In that case,  subsequent investigations have revealed that introducing a cosmological constant $\Lambda=3/L^2$ by considering the extended generalized uncertainty principle 
\beq
\Delta x \Delta p \sim \hbar +\alpha \frac{L_p^2 \Delta p^2}{\hbar} +\beta \frac{\hbar (\Delta x)^2}{L^2}
\eeq
will tame this discrepancy \cite{ong2}, while another possibility consists in taking into account the effects of the GUP not only in the number density but also in the equation of state $P=P(\tilde \varepsilon)$ of the matter constituting the white dwarfs interior \cite{rsos}. We will apply this latter way of thinking to our model in Sect. \ref{ss5}. 

\subsection{Exploring the Quadratic Correction in Eq. (\ref{cond2})}

When including the leading order corrections in the GUP*, (\ref{cond2}) becomes
\beq
\label{betac1}
\left[GM^2 -\frac{c\hbar}{2}\left(\frac{M}{m_e} \right)^{4/3}  \right]\frac{1}{R} -c \left(\frac{M\beta}{m_e} \right)^2 \left(\frac{\hbar}{R} \right)^3 +O(\beta^3)\sim 0 \,.
\eeq
First of all, we should note that possible corrections in $\beta$ arise only at second order (at which the sign of such parameter does not matter), which is consistent with the expansion of (\ref{solp}) when the square root comes with a negative sign:
\beq
\Delta p \approx \frac{\hbar}{2 \Delta x}\left[ 1+ \left(\frac{\hbar \beta}{2 \Delta x} \right)^2 +O(\beta^3) \right]\,.
\eeq
Next, the implicit $\beta$-corrected mass-radius relationship which follows from (\ref{betac1}) is 
\beq
\left[GM^2 -\frac{c\hbar}{2}\left(\frac{M}{m_e} \right)^{4/3}  \right]R^2 -c \left(\frac{M\beta}{m_e} \right)^2 \hbar^3 \sim 0\,,
\eeq
whose  physically relevant solutions (as $R$ should be positive)  depend on the sign of $\beta$ and read as\footnote{We stress that the Chandrasekhar limit should be identified {\it before} solving the GUP*-corrected equation for the gravitational equilibrium.}
\beq
\label{Rquadratic}
R \sim \pm \beta (2c\hbar^3)^{1/2} \left(\frac{M}{m_e}\right)^{1/3} [2G (M m_e^2)^{2/3} -c\hbar]^{-1/2}\,.
\eeq

\subsection{Exploring Some Routes for Restoring the Chandrasekhar Limit for the Plus Sign}

The equilibrium condition (\ref{eqrelativ}) holds assuming the validity of Newtonian gravity even in high pressure regimes. When investigating self-gravitating systems within the Jeans mass formalism, it has been argued that GUP*(\ref{newgup}) may affect the gravitational potential energy as \cite[Eq.(20)]{jeans}
\beq
|E_g^{corr}| \sim \frac{G M^2}{R}\left[1 +\frac{5 \ln (16 \pi R^2)}{144 R^2}\beta^2 \right]\,.
\eeq
However this is a quadratic correction in $\beta$ which, once the gravitational equilibrium condition $|E_g^{corr}| \sim E_k$ is imposed, cannot provide any mechanism for compensating the constant term $-cM/(\beta m_e)$.

By writing the potential energy with a Yukawa cut as
\beq
|E_g^{Yuk}| \sim \frac{GM^2}{R}(1+\delta e^{-\lambda R})\,,
\eeq
where $\delta$ quantifies the relative strength with respect to Newtonian gravity and $\lambda$ is the inverse of the force range, the gravitational force would experience a contribution which acts only on finite distances. This is related to massive gravity theories, in which the graviton particle mediating such interaction would no longer be massless, hence leading to a Yukawa potential.
Assuming again a small $\beta$ and now also a small $\lambda$, the condition for gravitational balance $|E_g^{Yuk}| \sim E_k$ reads as
\begin{eqnarray}
&& -\frac{cM}{\beta m_e} +\left[GM^2 (1+\delta)+\frac{c\hbar}{2}\left(\frac{M}{m_e} \right)^{4/3}  \right]\frac{1}{R} -GM^2\delta\lambda+O(\beta^2)+O(\lambda^2)\sim 0 \,.
\end{eqnarray}
Therefore a Yukawa-type potential allows us to compensate the previously identified constant term if we set the correction to the Newtonian gravitational force to actually oppose it with strength
\beq
\label{solalpha}
\delta = -\frac{c}{\beta GM m_e \lambda}<0\,,
\eeq 
where we have assumed the white dwarf mass to be a constant as from astrophysical observations. Then, at the lowest order in $\beta$, the gravitational equilibrium condition would become
\beq
GM^2 (1+\delta)+\frac{c\hbar}{2}\left(\frac{M}{m_e} \right)^{4/3}  \sim 0\,.
\eeq
A  solution to this equation consistent with the Chandrasekhar limit (\ref{cmass}) would require $\delta\sim -2$, and a negative $\delta$ is admissible from (\ref{solalpha}). Then, this implies a Yukawa cut at
\beq
\lambda \sim \frac{c}{2 \beta GM m_e}\,.
\eeq
Using $c/\beta   \gtrapprox  10^{19} \, {\rm GeV} \sim 1.60  \times 10^{9}$ J  \cite{highergup}, $G \sim 6.67 \times 10^{-11}$ N$\cdot$ m$^2$/kg$^2$, $m_e \sim 9.11 \times10^{-31} $ kg, $M \sim 2 \times 10^{30}$ kg, we can estimate   $\lambda \lesssim 0.7 \times 10^{19}$ m$^{-1}$.  For getting an intuitive idea of this order of magnitude, we can recall that the size of the dark matter halo in the Milky way is about $r_{dm} \sim 300 $ kpc $\sim  10^{22}$ m \cite{halo}, and therefore this type of correction in the gravitational potential would not affect the clustering properties of dark matter being relevant only on shorter length scales. More quantitatively, the value of a negative $\delta$ and of $1/\lambda $  we would need for restoring the Chandrasekhar limit have not been ruled out by recent analysis of galactic rotation curves \cite{iocco}. 

On the other hand, if we assume a post-Newtonian correction to the gravitational force that removes the infinity of the acceleration at very short distances and write 
\beq
E^{PPN} = \frac{GM^2}{(R+\epsilon)^\gamma}\,, \qquad \epsilon \approx 0\,, \qquad \gamma \approx 1\,,
\eeq
the gravitational equilibrium condition would provide 
\beq
-\frac{cM}{\beta m_e} +\left[GM^2 +\frac{c\hbar}{2}\left(\frac{M}{m_e} \right)^{4/3}  \right]\frac{1}{R} -\frac{\gamma + GM^2 (\gamma -1)(R-1) \ln R}{R^2}+O(\beta^2)+O(\epsilon^2)+O((\gamma-1)^2)\sim 0,
\eeq
without offering any possibility for compensating the previously mentioned constant term.

\section{Accounting for the GUP* corrections in the degenerate electron gas equation of state}
\label{ss5}

We will now compare the results for the mass-radius relations we have found thus far with the description of the degenerate electron gas, taking into account GUP*. Thus, similarly to what was done in \cite{rsos}, we will implement the equation of state for the degenerate electron gas we have found in Sect. \ref{sectgas} into the Tolman-Oppenheimer-Volkoff equations \cite{tov1,tov2}. They are given by the following system governing the radial evolution of the pressure and of the mass of the white dwarf:
\begin{eqnarray}
\label{tovm1}
\frac{\d P}{\d r} &=& -\frac{G ( \tilde \varepsilon +P) (M +4 \pi P r^3/c^2)}{c^2 r (r-2GM/c^2)}  \,, \\
\label{tovm2}
\frac{\d M}{\d r} &=& \frac{4 \pi \tilde \varepsilon r^2}{c^2}\,.
\end{eqnarray}
By introducing the dimensionless radial coordinate $\eta= (\alpha/(m_e c^2))^{1/3 } r$ and the dimensionless white dwarf mass $v=M/(m_e \pi)  $,  the TOV equations can be recast into the more convenient form
\begin{eqnarray}
\label{tov1}
\frac{\d \xi}{\d \eta} &=& \frac{q \sqrt{\xi^2 +1}(\sqrt{\xi^2 +1} + j-1) [ \tau \eta^3 + 5v]}{5 \xi (2 q v -\eta) \eta} \,; \\
\label{tov2}
\tau &=& \frac{5}{2}\ln(\sqrt{\xi^2 +1}+\xi) - \frac{(6 \tilde \beta \xi^4 -8 \tilde \beta \xi^2 -10 \xi^3 +16 \tilde \beta +15 \xi)\sqrt{\xi^2 +1}}{6} +\frac{8 \tilde \beta}{3}   \,, \\
\label{tov3}
\frac{\d v}{\d \eta} &=& -\eta^2 \left[ \frac{\ln(\sqrt{\xi^2 +1}+\xi)}{2} + \left( \frac{4 \tilde \beta \xi^4}{5} +\frac{4 \tilde \beta \xi^2}{15}  - \xi^3 -\frac{8 \tilde \beta}{15} -\frac{\xi}{2}\right) \sqrt{\xi^2 +1}  + (j-1) \left( \tilde \beta \xi -\frac{4}{3} \right) \xi^3   +\frac{8 \tilde \beta}{15} \right] \,,
\end{eqnarray}
where Eqs. (\ref{pressure})-(\ref{kineticb}) have been used, and in which we have defined $q := \pi G (m_e ^2 \alpha /c^8)^{1/3}$. We will now investigate the mass-radius relation in the low and high Fermi momentum limit separately.

In the non-relativistic regime for which $\xi \approx 0$ and considering $v \approx 0$, the approximated version of the TOV equations are
\beq
\label{approx1low}
\frac{\d \xi}{\d \eta} \approx -\frac{jq v}{\eta^2 \xi}  \,, \qquad \frac{\d v}{\d \eta} \approx \frac{4 j \eta^2 \xi^3}{3}     \,,
\eeq
in which, similar to the case in which GUP was considered in the degenerate electron gas equation of state \cite{rsos}, corrections beyond the usual Heisenberg's uncertainty relation does not affect the mass-radius relation of a white dwarf in the low momentum limit. In fact, searching for solution of the type $\xi \sim \eta ^s$, $ v \sim \eta ^t$, the previous system of differential equations becomes equivalent to
\beq
s -1 = t-2-s, \quad t-1 =2 +3s, \qquad \Rightarrow \qquad v \sim \eta^{-3} \qquad \Rightarrow \qquad R \sim M^{-\frac{1}{3}}        \,.
\eeq
Thus, also in this case general relativistic effects  resolve the mismatch between theory and observation  we identified in (\ref{issue}). The same result can be obtained in a more rigorous manner combining the two differential equations into a single  second order Lane-Emden equation as done in \cite{rsos}. Our result in the context of GUP* has the following intuitive explanation: in the low momentum approximation only the contribution of the electrons' rest masses matters in the white dwarf energy budget, and from (\ref{ngeneral}) we see that the GUP* correction are suppressed in this limit.

On the other hand, in the high momentum limit, for $\xi \to 1/\tilde \beta$, the TOV equation for the mass reduces to
\begin{eqnarray}
\frac{\d v}{\d \eta}  =  \eta^2 \theta_2    \,, \qquad
\theta_2 = \frac{ (16 \tilde \beta^4+7 \tilde \beta^2+6) \sqrt{1 + \tilde \beta^2}-16 \tilde \beta^5+10 (j-1) \tilde \beta -15 \tilde \beta^4 \ln \frac{\sqrt{1+\tilde \beta^2}+1}{\tilde \beta}}{30 \tilde \beta^4}\,,
\end{eqnarray}
from which the relation $v = \frac{\eta^3 \theta_2}{3} $, and $M \propto R^3$,  is easily obtained similarly to the scenario in which GUP was used \cite{rsos}. Next, approximating 
\beq
\label{mrhigh}
v \approx \frac{\eta^3}{15 \tilde \beta^4}
\eeq
for small $\tilde \beta $, we can write the TOV equation for the momentum as:
\begin{eqnarray}
\frac{\d \xi}{\d \eta} = \frac{q \sqrt{\xi^2 +1}(\sqrt{\xi^2 +1} + j-1) [3 \tau \tilde \beta^4 + 1] \eta}{15 \tilde \beta^4 \xi (s \eta^2 -1) }\,,
\end{eqnarray}
where $\tau$ is given in (\ref{tov2}) and $s := 2q/(15 \tilde \beta^4)$. Next, in the high-momentum limit we can approximate $\sqrt{\xi^2 +1}/\xi \approx 1$ and $\sqrt{\xi^2 +1} + j-1 \approx \xi +j$. Furthermore, by definition we have $q=\frac{\pi^{1/3} m_e^2 G}{\hbar c} \approx \tilde \beta^2$. Hence, we get
\beq
\frac{\d \xi}{\d \eta} \approx \frac{ (\xi + j) [3 \tau \tilde \beta^4 + 1] \eta}{15 \tilde \beta^2  (s \eta^2 -1) }\,, \qquad s=\frac{2}{15 \tilde \beta^2}\,.
\eeq
Now we note that at the same order of approximation we also have $3 \tau \tilde \beta^4 + 1 \approx -3 (\tilde \beta \xi)^5 + 5 (\tilde \beta \xi)^4 +\frac{5 (\tilde \beta \xi)^3 \tilde \beta^2}{2} -5 (\tilde \beta \xi)^2 \tilde \beta^2 \approx 2-\frac{5}{2 \xi^2}$ where in the last step we used $\tilde \beta =1/\xi$. Using this latter identification one more time, we finally arrive at
\beq
\frac{\d \xi}{\d \eta} \approx \frac{ (\xi + j) (4 \xi^2 -5) \eta}{30  (s \eta^2 -1) }\,, 
\eeq
which delivers the implicit solution 
\beq
\frac{\ln |s \eta^2 -1|}{60 s}  +\frac{1}{4 j^2 -5}   \left( \ln\frac{\sqrt{4 \xi^2 -5}}{\xi +j} +\frac{\sqrt{5}j}{5} \ln \frac{2 \sqrt{5} \xi +1}{2 \sqrt{5} \xi -1}  \right) +C      =0,
\eeq
with $C$ being an arbitrary constant of integration. Due to the smallness of $\tilde \beta$, in the high momentum limit we have
\beq
\frac{\tilde \beta^2}{8} \ln \frac{2 \eta^2}{15 \tilde \beta^2} +\frac{5 \ln(2) \xi -4j}{5(4 j^2 -5) \xi} +C \approx 0 \,,
\eeq
which provides
\beq
\label{etaxi}
\eta(\xi)=  \frac{\sqrt{30} \tilde \beta}{2} {\rm exp} \left( \frac{4 [4 j -20C j^2 \xi  +25 C \xi -5 \ln(2) \xi]  }{5 \tilde \beta^2 \xi (4 j^2 -5)}\right)\,.
\eeq
In the context of GUP, the Chandrasekhar limit was identified as $v_{\rm max} := \lim_{\xi \to \infty} v(\eta(\xi))$ \cite{rsos}. 
In the case under consideration, we set instead $v_{\rm max} := \lim_{\tilde \beta \to 0} v(\eta(\xi=1/\tilde \beta))$. Since $j=2 \times 10^4 \gg 1$, we get
\beq
v_{\rm max} \to \frac{1}{15 \tilde \beta^4} \left( \frac{\sqrt{30} \tilde \beta}{2}e^{-16 C /\tilde \beta^2} \right)^3\,.
\eeq
Now, we can note that if we choose the integration constant to be\footnote{As a consistency check we can note that $\lim_{\tilde \beta \to 0} C(\tilde \beta)=0$, which is indeed the same boundary condition adopted in \cite[Eqs.(3.16)-(3.17)]{rsos} when either GUP or the Heisenberg uncertainty relations are used. From the physical perspective, the condition we have chosen seems to arise from the fact that in the context of GUP* the Fermi momentum at the center of the star cannot achieve anymore arbitrarily large values being bounded above by $\tilde \beta^{-1}$. More quantitatively, recalling the definition of $\tilde \beta =m_e/M_p$ in terms of the mass of the electron and of the Planck mass, we have $C \sim -10^{-42}$ which is smaller by orders of magnitude  to both those energy scales and to that of $\tilde \beta$ itself. } $C \sim \frac{\tilde \beta^2}{16} \ln ( \sqrt{15/2} \tilde \beta^{2/3})$, we get $v_{\rm max} \sim \frac{1}{\tilde \beta^3}$, and $M_{\rm max} \sim \frac{M_p^3}{m_e^2}$ consistent with the Chandrasekhar limit.  Furthermore from (\ref{etaxi})  we get $\eta(\xi) \to \frac{\sqrt{30} \tilde \beta}{2} e^{-\frac{ \ln(2)}{\tilde \beta^2 j^2}} $ in the high-momentum limit, which is arbitrarily close to zero for $\tilde \beta \sim 0$.   This procedure is also consistent with the fact that only upper, not lower, limits on the parameter $\beta$  have so far been set  in literature (see \cite{nuclear} and references therein for various constraints arising from studies of the harmonic oscillator problem, hydrogen atom, equivalence principle, etc.), so that the modified theory should conceivably include  Chandrasekhar's one as an appropriate subcase. We should recall that in the GUP* model (\ref{newgup}) this is just an asymptotic state, being $\beta \neq 0$ by assumption; however our derivation shows in a mathematically transparent  way that for the subcase given by the usual Heisenberg uncertainty principle  the {\it infinity}-momentum solution of the TOV system corresponds to a vanishing radius configuration,  as indeed Chandrasekhar himself emphasized \cite{original2}. A finite momentum also prevents electrons' kinetic energy to become unphysically infinite, and consequently also the required counter-balancing gravitational energy  should not diverge anymore. Therefore, the ways GUP and GUP* affect the stellar structure are qualitatively different because in the former case,  when the electrons' momentum becomes (mathematically)  infinity, the white dwarf attains a  certain specific value of the radius \cite[Sect.3.1.2]{rsos}. On the other hand, a noteworthy similarity between the two models is that the commutation relation $[ \hat x , \, \hat p] \to \infty$ diverges in those scenarios, both when the electrons' momentum becomes (mathematically) infinity in GUP and when it approaches $ \sim 1/\beta$ in GUP*.   



We need to comment that in our procedure the Chandrasekhar limit does \emph{not} arise as the zeroth order term in a series expansion in $\tilde \beta$ but this parameter  is actually the one setting the scale of the Chandrasekhar mass; therefore,  the subcase of the standard Heisenberg principle should be handled setting $\tilde \beta=0$ in the TOV equations before any further manipulations because in the context of GUP* this parameter represents the energy scale of  the electrons' momentum in the ultrarelativistic regime. The purpose of our work is not to try to account for exotic astrophysical observations which may challenge the Chandrasekhar limit by adopting (\ref{newgup}) because it would anyway remain mysterious why some stellar configurations rather than others should obey some specific uncertainty principle. Here we wanted to prove that an appropriate mathematical formalism which allows us to identify the Chandrasekhar limit, should we adopt GUP*, exists in a quite conservative manner, for example without invoking any dependence of the spatial uncertainty $\Delta x$ on the number of species of particles constituting the white dwarf \cite{NL1,NL2}, nor introducing some non-commutativity in the spacetime geometry \cite{NL3}, but keeping the TOV equations as formulated in general relativity. Therefore, according to us the higher-order uncertainty principle with maximum momentum proposed in \cite{highergup} by some other authors has at least passed this astrophysical test and it might be further applied in future investigations. 

Although it might be tempting to claim that, likewise, the Oppenheimer-Volkoff limit for neutron stars would emerge {\it mutatis mutandis} by replacing the electron mass with the neutron mass, we warn that both the equations of state in (\ref{pressureb}), (\ref{entot}), as well as in \cite{rsos}, do not seem physically appropriate for the description of nuclear matter. In fact, in both scenarios the pressure vanishes when the energy density does\footnote{ The equations of state $P=P(\tilde \varepsilon)$ are given in a parametric form where the parameter is $\xi$. It is easy to check that for $\xi=0$ we have both $P=0$ and $\tilde \varepsilon =0$ implying that the pressure is zero when the energy density vanishes.}. However, a key property of the equation of state for nuclear matter is that the pressure is nonzero even when the energy density is. This is a consequence of the role played by the {\it bag constant} \cite{bag1,bag2}. Therefore it seems necessary to look for a completely different route for encoding GUP effects into a realistic modeling of neutron stars.    

We will now briefly address the stability properties of the configuration described by the TOV equations in which  we have encoded the GUP* corrections in the description of the degenerate electron gas. 
The adiabatic speed of sound, which sets the speed at which perturbations propagate, within the degenerate electron gas is
\beq
\label{velocity}
c_s^2 = \frac{\partial P}{\partial \tilde \varepsilon}= \frac{ \xi^2 (4 -3 \tilde \beta \xi)}{12 [(j-1) \sqrt{1+ \xi^2}+1+ \xi^2](1-\tilde \beta \xi)}\,,
\eeq
where Eqs. (\ref{pressure}), (\ref{kineticb}), (\ref{entot}) have been used, and the chain rule for derivatives has been applied via the auxiliary variable $\xi$. Recall that $j>0$, and notice that the argument of the square root is larger than 1. For a positive $\tilde \beta$, which is the case of interest in Doubly Special Relativity and for which $\xi \leqslant 1/\tilde \beta$, it can be concluded that $c_s^2>0$ because all its factors are positive (this result can also be obtained by noticing from Fig. \ref{fig1} that the energy density $\tilde \varepsilon=\tilde \varepsilon(\xi)$ is an increasing function of the electrons' momentum, implying that its inverse $\xi=\xi(\tilde\varepsilon)$ is increasing as well, and finally $P=P(\tilde\varepsilon)$ is increasing by composition of increasing functions), and therefore stability by the Le Ch\^atelier principle can be established  \cite{chatelier}. Moreover, in units of $\tilde \beta =1$ we have $j = m_u \mu_e /M_p \gg  1$ by many orders of magnitude, and the condition $c_s^2 <1$ becomes equivalent to $12 (j-1) \sqrt{1+\xi^2}(1-\xi) \geqslant 9 \xi^3 -8 \xi^2 +12 \xi -12$,  which holds due to the fact that, as required by $\xi \leqslant 1$, the leading order term in the LHS is $j$, which is larger than the leading order term in the RHS.

\begin{figure}[h!]
	\begin{center}$
		\begin{array}{cccc}
		\includegraphics[width=85 mm]{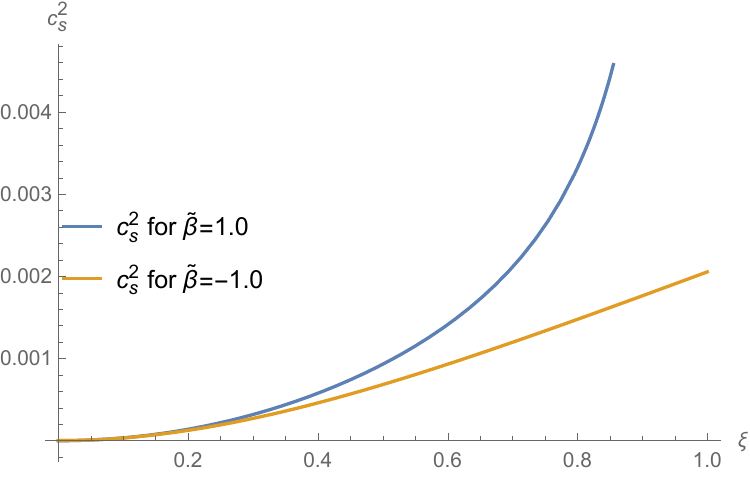}&&
		\includegraphics[width=85 mm]{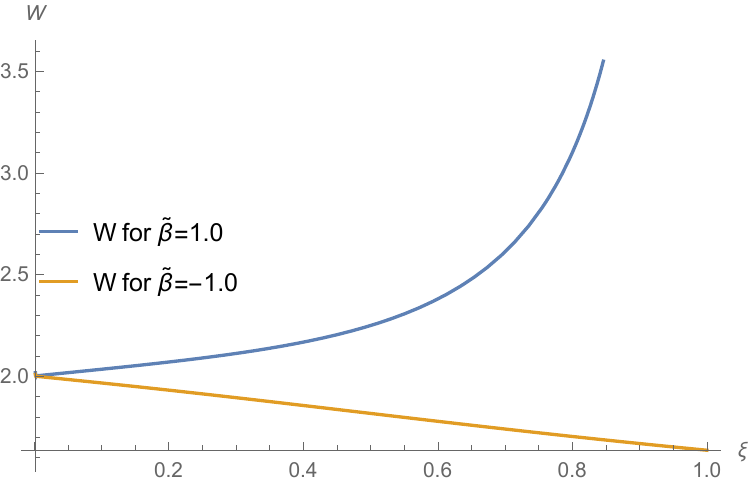}\\
		(a)    && (b)
		\end{array}$
	\end{center}
	\caption{In panel (a) we display the adiabatic speed of sound $c_s^2$ within the degenerate electron gas obeying GUP* computed in (\ref{velocity}) in which we  have set $j=100$, while in panel $b$ the function $W$ from (\ref{eqW}). We can note that the former quantity is bounded between 0 and 1, while the latter is positive confirming the stability of the white dwarf. We have considered the cases of $\tilde \beta=\pm 1$ with the top curves referring to the positive ones in both panels. }
	\label{fig2}
\end{figure}

Our graphical analysis reported in panel (a) of Fig. (\ref{fig2}) -- in which for completeness we consider  also the case of negative $\tilde \beta$ -- confirms our results.
Then, the stability analysis  requires us to also check that the increase of perturbations should actually require electrons to be provided with more energy from external sources. Since this would affect the speed of electrons but not their rest masses we fix $j=0$, and from the TOV equations (\ref{tov1})-(\ref{tov2})-(\ref{tov3}) and the expression for the electrons kinetic energy (\ref{kineticb})
we can compute 
\beq
\frac{\partial v}{\partial \epsilon}= \frac{4 \left[\frac{5 \ln(\sqrt{1+\xi^2} +\xi)}{8} +\left(\tilde \beta \xi^4 +\frac{\tilde \beta \xi^2}{3} - \frac{5 \xi^3}{4} -\frac{2 \tilde \beta}{3} -\frac{5 \xi}{8} \right) \sqrt{1+\xi^2} -\frac{5 \tilde \beta \xi^4}{4} +\frac{5 \xi^3}{3} +\frac{2 \tilde \beta}{3}\right] (2qv -\eta) \eta^3}{(1-\sqrt{1 +\xi^2}) \left[ \frac{5 \eta^3 \ln(\sqrt{1+\xi^2} +\xi)}{2} - \eta^3 \left(\tilde \beta \xi^4 -\frac{4 \tilde \beta \xi^2}{3} - \frac{5 \xi^3}{3} +\frac{8 \tilde \beta}{3} +\frac{5 \xi}{2}  \right) \sqrt{1+\xi^2} +\frac{8 \tilde \beta \eta^3}{3} +5v\right] \xi (\xi^2 + 1 -\sqrt{1+ \xi^2}) \alpha (1-\tilde \beta \xi)}\,,
\eeq
which for objects of small radius can be approximated as
\beq
\frac{\partial v}{\partial \epsilon}= \frac{8 \left[\frac{5 \ln(\sqrt{1+\xi^2} +\xi)}{8}  +\left(\tilde \beta \xi^4 +\frac{\tilde \beta \xi^2}{3} - \frac{5 \xi^3}{4} -\frac{2 \tilde \beta}{3} -\frac{5 \xi}{8} \right) \sqrt{1+\xi^2} -\frac{5 \tilde \beta \xi^4}{4}  + \frac{5 \xi^3}{3} + \frac{2 \tilde \beta}{3}\right] \eta^3}{5 (1-\sqrt{1+ \xi^2}) \xi( \xi^2 +1 -\sqrt{1+\xi^2})q \alpha (1-\tilde \beta \xi)} +O(\eta^4)\,.
\eeq
In panel (b) of Fig. (\ref{fig2}) we confirm that the function
\beq
\label{eqW}
W = \frac{\frac{5 \ln(\sqrt{1+\xi^2} +\xi)}{8}  +\left(\tilde \beta \xi^4 +\frac{\tilde \beta \xi^2}{3} - \frac{5 \xi^3}{4} -\frac{2 \tilde \beta}{3} -\frac{5 \xi}{8} \right) \sqrt{1+\xi^2} -\frac{5 \tilde \beta \xi^4}{4}  + \frac{5 \xi^3}{3} + \frac{2 \tilde \beta}{3} }{ (1-\sqrt{1+ \xi^2}) \xi( \xi^2 +1 -\sqrt{1+\xi^2})  (1-\tilde \beta \xi)}
\eeq
is positive, and therefore the stellar configuration is stable because $\frac{\partial M}{\partial \epsilon}>0$ meaning that the electrons need to be supplied with energy from the exterior for perturbations to develop \cite[pages 304-308]{wein}.  

Last but not least, our configuration fulfills also the Buchdahl criterion for stability \cite{buch} because, as a direct consequence of the existence of a maximum value for the electrons' momentum, the pressure (\ref{pressure}) cannot diverge at any point of the star.  Having a possibly more physically realistic always-finite pressure is a second remarkable difference compared to the GUP   and Heisenberg scenarios,  in those cases the pressure diverges in correspondence of the Chandrasekhar limit \cite{rsos}.  In particular, by using (\ref{largepressure}) we can estimate that the maximum value of the pressure, which corresponds to a white dwarf with Chandrasekhar mass, e.g. to the high-momentum regime, is $P \approx  \frac{\alpha}{30 \tilde \beta^4} \approx \frac{c^7}{30 \pi^2 \hbar G^2 }$.  Although the value  of the momentum of the electrons is bounded above, it can nevertheless be high enough to provide the amount of kinetic energy required for electron capture (white dwarfs are globally electrically neutral, and so they need to contain also protons),  compensating for the difference in the rest mass between a proton and a neutron. This then leads to the conversion of an electron and a proton into a neutron and an electric neutrino, and hence GUP* does not forbid the explosion of white dwarfs into supernovae \cite{garf}.

\section{Modifications of gravity and  temperature effects}
\label{ss6a}

In these two subsections separately, we will explore some effects within the TOV equations implied by some modifications of the gravity sector and  arising from a non-zero temperature for the electrons gas. In fact, a specific set of hydrostatic equations governing the  equilibrium in the interior of a star, but without affecting its exterior gravitational field,  is a prediction of some one-parameter beyond-general-relativity theories not yet ruled out by gravitational wave observations \cite{ref14,ref15}. In \cite{jcapref}, a new equation for the radial evolution of the pressure has been considered for some theories with a non-Brans-Dicke coupling between scalar field and curvature, and nevertheless free from  Ostrogradsky instabilities; we will assess the role of GUP* within this modified TOV equation in what follow. On the other hand, if the electrons' gas is not anymore degenerate  we need to account for   its finite temperature when computing its TOV pressure and energy density via the Fermi-Dirac statistics: we will claim that actually the deviation from the Heisenberg uncertainty principle we have investigated in this paper  behaves as a temperature effect.

\subsection{Modifications to  the radial evolution of the pressure }

By following \cite{jcapref}, we keep the usual TOV equation for the mass (\ref{tovm2}), but we replace the equation for the pressure (\ref{tovm1})  with
\beq
\frac{\d P}{\d r}= -\frac{G}{c^2} \left( \frac{M(r)}{r^2} - \frac{Y}{4} \frac{d^2 M(r)}{\d r^2}\right)  \tilde \varepsilon(r)\,.
\eeq
This equation follows from the modified Poisson equation for the gravitational field $\Phi$
\beq
\nabla^2 \Phi= \frac{4 \pi G}{c^2} \tilde \varepsilon +\frac{GY}{4 c^2}  \nabla^2 \left( \frac{d M}{d r} \right)\,,
\eeq
where $Y$ is a parameter quantifying the breaking of the Vainshtein screening mechanism, and consequently the deviation from standard gravity \cite{ref14,ref18,ref19}. By using again the chain rule with the electrons' momentum $\xi$ as the auxiliary variable and Eq.(\ref{tovm2}), we obtain
\beq
\frac{\d \xi}{\d r} =  \frac{G ( 2 \pi Y r^3 \tilde \varepsilon -c^2 M ) \tilde \varepsilon}{(rc^2)^2 \frac{\d P}{\d \xi} - \pi G Y r^4 \tilde \varepsilon \frac{\d  {\tilde\varepsilon}}{\d \xi}}\,,
\eeq
which explicitly turns out to be
\begin{eqnarray}
\frac{\d \xi}{\d r} &=& \frac{4  G \sqrt{\xi^2 +1} \cdot  {\mathcal S}_1 \cdot ( 5c^2M - \pi \alpha Y {\mathcal S}_1 r^3 )}{ 25 r^2 {\mathcal S}_2 }\,, \\
{\mathcal S}_1 &=& \frac{5}{8} \ln(\sqrt{\xi^2+1}+\xi)+ \left(\tilde \beta \xi^4+\frac{\tilde \beta}{3} \xi^2-\frac{5}{4} \xi^3-\frac{2 \tilde \beta}{3}-\frac{5}{8}\xi \right)\sqrt{\xi^2+1}+\frac{5 \tilde \beta}{4}(j-1)\xi^4+\frac{5}{3}(1-j)\xi^3+\frac{2 \tilde \beta}{3} \,, \\
{\mathcal S}_2 &=& \frac{5}{4}(\tilde \beta \xi-1)G \alpha Y [(j-1)+\sqrt{\xi^2+1}] \sqrt{\xi^2+1}r^2 \,{\mathcal S}_1 +  c^4 \xi^4 \left(  \tilde \beta \xi - \frac{4}{3} \right)\,.
\end{eqnarray}	
By adopting the same re-scaled variables and parameters as in (\ref{tov1}), this latter equation can be recast as
\begin{eqnarray}
\label{MMM1}
\frac{\d \xi}{\d \eta} &=& \frac{q \sqrt{\xi^2 +1} \cdot   {\mathcal S}_1 \cdot ( 5v-  Y  {\mathcal S}_1 \eta^3 )}{ 25 \eta^2 \tilde{\mathcal S}_2 }\,, \\
\tilde {\mathcal S}_2 &=& \frac{5}{4}(\tilde \beta \xi-1) Y [(j-1)+\sqrt{\xi^2+1}] \sqrt{\xi^2+1}\eta^2 \,{\mathcal S}_1 +   \xi^4 \left(  \tilde \beta \xi - \frac{4}{3} \right)\,.
\end{eqnarray}
For low momentum $ \xi \approx 0$, at the lowest order of approximation, eq.(\ref{approx1low}) is re-obtained and therefore the parameter $Y$ plays a negligible role in this regime\footnote{More quantitatively, at the denominator we would have $\xi \to \xi(1+Y\xi)$ with the second term smaller by orders of magnitude than the first.}. Furthermore, the correction proportional to $Y$ does not affect the high momentum regime either. We can prove this by recalling that the scaling for the mass $v \sim 1/\tilde \beta^4$ from (\ref{mrhigh}) still holds being its corresponding TOV equation unaffected by the modifications in the Poisson equation; then, in the numerator of (\ref{MMM1}) we would have ${\mathcal S}_1 \sim \xi^3 \sim 1/\tilde \beta^3$ by recalling that in the high momentum regime $\xi \to 1/\tilde \beta$, and thus the correction plays a second order role; likewise at the denominator, the term proportional to $Y$ goes as $\sim \xi^2$ which is suppressed compared to $\xi^4$ in the high momentum regime. Thus, the lack of the Vainshtein screening does not influence our qualitative results in these two limiting cases (it may nevertheless play a role in shaping the mass-radius relationship  for intermediate values of $\xi$). Our conclusion seems indeed in agreement with that of \cite{valoriY} which set stringent constraints on the possible values of $Y$ in the context  of white dwarfs.

\subsection{Temperature effects}

The Fermi-Dirac distribution reads as
\beq
\label{FM}
f_T(p) =\frac{1}{1+e^{\frac{E_p -m_ec^2 - \mu}{k_B T}}}\,,
\eeq
where $E_p$ is the relativistic energy of a single electron introduced in (\ref{defenergy}), $\mu$ the chemical potential, $k_B$ the Boltzmann constant and $T$ the absolute temperature. When dealing with the Fermi-Dirac statistics, that the microscopic constituents of white dwarfs obey to, thermodynamical quantities can be written as functions of the  Fermi-Dirac integral  \cite{integrals1}:
\beq
F_k(\eta, \alpha) := \int_0^\infty \frac{t^k \sqrt{1 +(\alpha/2)t}}{1+e^{t-\eta}}dt \,.
\eeq
We think however that it might be more instructive to outline the temperature effects as follow. The Fermi-Dirac distribution (\ref{FM}) further restricts the available phase space; this means we should expect the density $ n$ to be smaller than the one computed for a degenerate configuration and,  from $m_e  n  \sim M/V$,  this means that the star is less compact.

More interestingly, we recall that GUP* restricts as well the available phase space, and also removes the contributions to the density from the electrons' at infinite momentum, essentially behaving as a finite temperature effect. The analogy actually looks like deeper; by considering the asymptotic
\beq
f_T(p) \approx \frac{1}{2} - \frac{\sqrt{(pc)^2 + (m_ec^2)^2}-\mu}{k_B T}  \,, \qquad  {\rm for}\, \, T\gg 1 \,,
\eeq 
we note, first of all, that the leading order  correction is linear in the momentum $p$, as predicted by GUP*. Considering further a situation in which the kinetic energy dominates over the rest energy and the chemical potential, we can {\it conceptually} identify
\beq
\beta \sim \frac{c}{4 k_B T}\,.
\eeq
We stress that this identification is only conceptual: while $\beta$ behaves like a finite temperature effect, the modification to the uncertainty principle in (\ref{newgup}) does not arise from  the temperature, for otherwise the value of $\beta$ would change as $T$ is varied.

\section{Discussion}
\label{ss6}

The theory of Doubly Special Relativity predicts the existence of a maximum momentum and a minimum length. This requirement should be realized via an appropriate modified uncertainty relation GUP* beyond the Heisenberg model which then allows explicit computations of various observable quantities in different physical frameworks. One explicit proposal of a new GUP* within this theory has been put forward in \cite{highergup} which constitutes the basic assumption of this paper that we have introduced in (\ref{newgup}).  From a mathematical perspective, this actually constitutes a family of four uncertainty relations due to the possible combinations of the sign of the free parameters quantifying the deviations from the standard momentum-position commutation relation (only one of which is consistent with the Doubly Special Relativity requirements though) and of a certain sign in front of a square root when solving for the momentum. Therefore, the different applicability of these four possibilities calls for a quantitative assessment. Along the same line of thinking it was clarified in  recent literature  that specific choices of the sign of the free parameter in GUP  lead to completely different cosmological dynamics with  either a singularity or a bounce occurring \cite{gabriele}.

In this paper, we have tried to describe a white dwarf in gravitational equilibrium in light of GUP*. First of all, we have applied some qualitative arguments and obtained that in the non-relativistic regime the Chandrasekhar mass is no longer reached as a limiting condition, and that the radius of the white dwarf is an increasing function of the mass; similarly to the context in which the GUP (\ref{gupeq}) was considered \cite{rsos}, this inconsistency with astrophysical observations is resolved when the Tolman-Oppenheimer-Volkoff equations are used. On the other hand, in the relativistic regime the Chandrasekhar mass can be  identified already from a qualitative analysis, but only if a specific sign in front of the previously mentioned square root in (\ref{solp}) is considered; in the other case restoring this limit is not a trivial task as we have discussed with reference to Yukawa and Post-Newtonian corrections to the gravitational force, and indeed this scenario may be simply excluded as it seems to also require a not-so-natural corrections to the relationship between horizon radius and mass of a black hole in order to avoid the white dwarf from degenerating into that configuration. 

The requirement of a maximum momentum at high energy leads to a constant value for the white dwarf mass from the TOV equations. As a side result we have computed also the pressure and energy of a degenerate electron gas as we needed them for integrating the TOV equations; specifically we have confirmed that both of these quantities are positive. In other words GUP* is the only ``new physics'' considered here. This should be contrast with exotic matter fields found in the context of other applications of modified uncertainty relations, which have been invoked for characterizing the nature of dark energy and/or the inflationary dynamics of the Universe \cite{dark1,dark2,dark3,dark4}. Since we only analyzed non-exotic matter, this part of our results can in principle be relevant also in condensed matter physics for the description of the conductivity properties of various materials. In addition, the equation of state of the degenerate electron gas can as well be implemented in the field equations arising from the Lema\^itre-Tolman-Bondi metric for tracking the gravitational collapse of material leading to the formation of the white dwarf; this can constitute the subject of future research since in the current paper we have assumed the conditions for gravitational equilibrium to be in place as in the TOV framework and then confirmed explicitly the stability of our configuration.

Finally, let us return to a crucial difference between GUP and GUP*: the fact that in the \emph{classical} limit ($\hbar \to 0$) the latter recovers $\Delta x\Delta p \sim 0$, just like the usual Heisenberg's uncertainty principle, whereas the former (with positive GUP parameter) does \emph{not}. In fact, the classical limit of GUP is equivalent to $\Delta x \sim {G\Delta p}/{2c^3}$. This has an obvious interpretation, in the classical limit but when gravitational effect is dominant, a particle of mass $m$ cannot be localized to within an uncertainty $\Delta x$. In the language of Heisenberg's microscope, if one focuses a beam of light with small enough wavelength (high enough energy) in order to localize the particle, an event horizon is formed and so we cannot locate the particle to within the horizon scale $\sim Gm/c^2$. The change in the momentum of the particle is $\gamma m v$, which for $v$ close to $c$ but still bounded away from $c$, is $O(1) mc$ (e.g., at $v=0.9 c$, $\gamma m v\approx 2 mc$). Thus we see that $\Delta x \sim {G\Delta p}/{2c^3}$ does lead to $\Delta x \sim Gm/c^2$. This reasoning can be reformulated in the following way \cite[Sect.2]{iorio}:  quantum fluctuations at Planck scale can lead to the formation of (virtual) micro-black holes and the measurement process is making  the gravitational radius to become wider up to the same width of $\Delta x$ when the critical value of a minimum observable region given by Planck length is reached \cite{scardigli}. This argument also suggests that the modifications to the Heisenberg principle are sensitive to both the number of spatial dimensions and to the specific black hole solution considered (e.g. Schwarzschild in this case) via the expression of the black hole radius. In other words, GUP combines the effect of Heisenberg's uncertainty principle with that of the uncertainty due to horizon formation in the presence of gravity, therefore in the classical limit it still reduces to the classical effect involving gravity, and not to the flat spacetime limit (unless $G \to 0$). On the other hand, the classical limit of GUP* reduces to that of flat spacetime.  

From this perspective it is not surprising that the Chandrasekhar limit for GUP* corresponds to a zero radius star (just like when the original Heisenberg's uncertainty principle is used), as there is no cutoff that corresponds to horizon formation. We emphasize that this is not a shortcoming: as in the original Chandrasekhar limit, it is an asymptotic state that is not attainable, as complicated stellar physics often lead to instabilities that result in most \emph{stable} white dwarfs being bounded away from the Chandrasekhar limit.  One could argue that the fact that the stellar size is zero in the original Chandrasekhar limit is the sign that the theory ignores general relativistic effect  in the formulation of the uncertainty principle, and thus the fact that GUP leads to a nonzero radius is preferred over GUP*. In other words, that the flat spacetime limit is not recovered unless $G \to 0$ could therefore be a desired feature. On the other hand, white dwarfs depend on both $G$ and $\hbar$ being nonzero,  so a classical limit with nonzero $G$ seems irrelevant.  

This actually raises an interesting conceptual question for the generalized/modified uncertainty principle community. Often it is useful to have a guiding principle or criterion that one can rely on when attempting to study quantum gravity phenomenology. Should we require that the classical limit recover the flat spacetime limit or the horizon formation criterion? Should a quantum uncertainty be put on equal footing with a classical one?

\begin{acknowledgments}
DG is a member of the GNFM working group of the Italian INDAM. DG acknowledges economical support from the start-up plan of Jiangsu University of Science and Technology.  YCO thanks the National Natural Science Foundation of China (No.11922508) for funding support. 
\end{acknowledgments}

{}


\begin{thebibliography}{99}
	
	
\bibitem{gupi1}
Gabriele Veneziano, ``A Stringy Nature Needs Just Two Constants'', {\hypersetup{urlcolor=vividviolet}\href{https://iopscience.iop.org/article/10.1209/0295-5075/2/3/006}{Europhys. Lett. \textbf{2} (1986) 199}}.
		
\bibitem{gupi2}
David J. Gross, Paul F. Mende, ``The High-Energy Behavior of String Scattering Amplitudes'', {\hypersetup{urlcolor=vividviolet}\href{https://www.sciencedirect.com/science/article/abs/pii/0370269387903558?via\%3Dihub}{Phys. Lett. B \textbf{197} (1987) 129}}.
		
\bibitem{gupi3}
David J. Gross, Paul F. Mende, ``String Theory Beyond the Planck Scale'', {\hypersetup{urlcolor=vividviolet}\href{https://www.sciencedirect.com/science/article/abs/pii/0550321388903902?via\%3Dihub}{Nucl. Phys. B \textbf{303} (1988) 407}}.
		
		
\bibitem{gupi4}
Michele Maggiore, ``A Generalized Uncertainty Principle in Quantum Gravity'', {\hypersetup{urlcolor=vividviolet}\href{https://www.sciencedirect.com/science/article/abs/pii/0370269393914018?via\%3Dihub}{Phys. Lett. B \textbf{304} (1993) 65}}, \href{https://arxiv.org/abs/hep-th/9301067}{[arXiv:hep-th/9301067]}.
		
		
\bibitem{gupi5}
Abdel Nasser Tawfik, Abdel Magied Diab, ``Generalized Uncertainty Principle: Approaches
and Applications'', {\hypersetup{urlcolor=vividviolet}\href{https://www.worldscientific.com/doi/abs/10.1142/S0218271814300250}{Int. Jour. Mod. Phys. D \textbf{23} (2014) 1430025}}, \href{https://arxiv.org/abs/1410.0206}{[arXiv:gr-qc/1410.0206]}.

\bibitem{paff}
Kenichi Konishi, Giampiero Paffuti, Paolo Provero, ``Minimum Physical Length and the Generalized Uncertainty Principle in String Theory'', {\hypersetup{urlcolor=vividviolet}\href{https://www.sciencedirect.com/science/article/abs/pii/0370269390919274?via\%3Dihub}{Phys. Lett. B \textbf{234} (1990) 276}}.

\bibitem{gupi6}
Ronald J. Adler, Pisin Chen, David I. Santiago, ``The Generalized
Uncertainty Principle and Black Hole Remnants'', {\hypersetup{urlcolor=vividviolet}\href{https://link.springer.com/article/10.1023\%2FA\%3A1015281430411}{Gen. Rel. Grav. \textbf{33} (2001) 2101}}, \href{https://arxiv.org/abs/gr-qc/0106080}{[arXiv:gr-qc/0106080]}.




\bibitem{gupi7}
Pisin Chen, Ronald J. Adler, ``Black Hole Remnants and Dark Matter'', {\hypersetup{urlcolor=vividviolet}\href{https://www.sciencedirect.com/science/article/abs/pii/S0920563203020887?via\%3Dihub}{ Nucl. Phys. Proc. Suppl. \textbf{124} (2003) 103}}, \href{https://arxiv.org/abs/gr-qc/0205106}{[arXiv:gr-qc/0205106]}.


\bibitem{baryo}
Saurya Das, Mitja Fridman, Gaetano Lambiase, Elias C. Vagenas, ``Baryon Asymmetry from the Generalized Uncertainty Principle'', {\hypersetup{urlcolor=vividviolet}\href{https://www.sciencedirect.com/science/article/pii/S0370269321007814?via\%3Dihub}{ Phys. Lett. B \textbf{824} (2022) 136841}}, \href{https://arxiv.org/abs/2107.02077}{[arXiv:gr-qc/2107.02077]}.

\bibitem{gw}
Mohamed Moussa, Homa Shababi, Ahmed Farag Ali, ``Generalized Uncertainty Principle and Stochastic Gravitational Wave Background Spectrum'', {\hypersetup{urlcolor=vividviolet}\href{https://www.sciencedirect.com/science/article/pii/S0370269321000113?via\%3Dihub}{ Phys. Lett. B \textbf{814} (2021) 136071}}, \href{https://arxiv.org/abs/2101.04747}{[arXiv:gr-qc/2101.04747]}.


\bibitem{amelino0}
Giovanni Amelino-Camelia, Jerzy Kowalski-Glikman, Gianluca Mandanici, Andrea Procaccini, ``Phenomenology of Doubly Special Relativity'', {\hypersetup{urlcolor=vividviolet}\href{https://www.worldscientific.com/doi/abs/10.1142/S0217751X05028569}{Int. Jour. Mod. Phys. A \textbf{20} (2005) 6007}}, \href{https://arxiv.org/abs/gr-qc/0312124}{[arXiv:gr-qc/0312124]}.
		
\bibitem{pedram1}
Pouria Pedram, ``A Higher Order GUP with Minimal Length Uncertainty and Maximal Momentum'', {\hypersetup{urlcolor=vividviolet}\href{https://www.sciencedirect.com/science/article/pii/S0370269312007459?via\%3Dihub}{Phys. Lett. B \textbf{714} (2012) 317}}, \href{https://arxiv.org/abs/1110.2999}{[arXiv:hep-th/1110.2999]}.
		
		
\bibitem{pedram2}
Pouria Pedram, ``A Higher Order GUP with Minimal Length Uncertainty and Maximal Momentum II: Applications'', {\hypersetup{urlcolor=vividviolet}\href{https://www.sciencedirect.com/science/article/pii/S0370269312011252?via\%3Dihub}{Phys. Lett. B \textbf{718} (2012) 638}}, \href{https://arxiv.org/abs/1210.5334}{[arXiv:hep-th/1210.5334]}.



\bibitem{highergup}
Won Sang Chung, Hassan Hassanabadi, ``A New Higher Order GUP: One Dimensional Quantum System'', {\hypersetup{urlcolor=vividviolet}\href{https://link.springer.com/article/10.1140\%2Fepjc\%2Fs10052-019-6718-3}{Eur. Phys. J. C \textbf{79} (2019) 213}}.
		
		
\bibitem{weincc}
Steven Weinberg, ``The Cosmological Constant Problem'', {\hypersetup{urlcolor=vividviolet}\href{https://journals.aps.org/rmp/abstract/10.1103/RevModPhys.61.1}{Rev. Mod. Phys.  \textbf{61} (1989) 1}}.		
		
\bibitem{ong1}
Yen Chin Ong, ``Generalized Uncertainty Principle, Black Holes, and White Dwarfs: A Tale of Two Infinities'', {\hypersetup{urlcolor=vividviolet}\href{https://iopscience.iop.org/article/10.1088/1475-7516/2018/09/015}{JCAP \textbf{09} (2018) 015}}, \href{https://arxiv.org/abs/1804.05176}{[arXiv:gr-qc/1804.05176]}.

\bibitem{ong2}
Yen Chin Ong, Yuan Yao, ``Generalized Uncertainty Principle and White Dwarfs Redux: How Cosmological Constant Protects Chandrasekhar Limit'', {\hypersetup{urlcolor=vividviolet}\href{https://journals.aps.org/prd/abstract/10.1103/PhysRevD.98.126018}{Phys. Rev. D \textbf{98} (2018)  126018}}, \href{https://arxiv.org/abs/1809.06348}{[arXiv:gr-qc/1809.06348]}.

\bibitem{annals}
Reza Rashidi,  ``Generalized Uncertainty Principle and the
Maximum Mass of Ideal White Dwarfs'', {\hypersetup{urlcolor=vividviolet}\href{https://www.sciencedirect.com/science/article/abs/pii/S0003491616301890?via\%3Dihub}{Annals Phys.  \textbf{374} (2016)  434}}, \href{https://arxiv.org/abs/1512.06356}{[arXiv:gr-qc/1512.06356]}.		
		
\bibitem{rsos}
Arun Mathew, Malay Kumar Nandy,   ``Existence of Chandrasekhar's limit in generalized uncertainty white dwarfs'', {\hypersetup{urlcolor=vividviolet}\href{https://royalsocietypublishing.org/doi/10.1098/rsos.210301}{ R. Soc. Open Sci.  \textbf{8} (2021)  210301}}, \href{https://arxiv.org/abs/2002.08360}{[arXiv:gr-qc/2002.08360]}. 		
	
\bibitem{original1}
Subrahmanyan Chandrasekhar, Edward Arthur Milne,  ``The Highly Collapsed Configurations of a Stellar Mass'', {\hypersetup{urlcolor=vividviolet}\href{https://academic.oup.com/mnras/article/91/5/456/985147}{ 	MNRAS  \textbf{91} (1931) 456}}.



\bibitem{original2}
Subrahmanyan Chandrasekhar,   ``The Highly Collapsed Configurations of a Stellar Mass. (Second Paper.)'', {\hypersetup{urlcolor=vividviolet}\href{https://academic.oup.com/mnras/article/95/3/207/991812}{ 	MNRAS  \textbf{95} (1935) 207}}.	

\bibitem{landau}
Lev D. Landau, ``On the Theory of Stars'', Phys. Z. Sowjetunion \textbf{1} (1932) 285.

\bibitem{stoner}
Edmund C. Stoner,   ``The Limiting Density in White Dwarf Stars'', {\hypersetup{urlcolor=vividviolet}\href{https://www.tandfonline.com/doi/abs/10.1080/14786440108564713}{ Philosophical Magazine  \textbf{7} (1929)  63}}.

	
\bibitem{tov1}
Richard C. Tolman,   ``Static Solutions of Einstein's Field Equations for Spheres of Fluid'', {\hypersetup{urlcolor=vividviolet}\href{https://journals.aps.org/pr/abstract/10.1103/PhysRev.55.364}{ Phys. Rev.  \textbf{55} (1939)  364}}.


\bibitem{tov2}
J. Robert Oppenheimer,  George Volkoff,   ``On Massive Neutron Cores'', {\hypersetup{urlcolor=vividviolet}\href{https://journals.aps.org/pr/abstract/10.1103/PhysRev.55.374}{ Phys. Rev.   \textbf{55} (1939)  374}}.			
		
		
\bibitem{annu1}
Wolfgang Hillebrandt,  Jens C. Niemeyer, ``Type Ia Supernova Explosion Models'', {\hypersetup{urlcolor=vividviolet}\href{https://www.annualreviews.org/doi/10.1146/annurev.astro.38.1.191}{Ann. Rev. Astron. Astrophys. \textbf{38} (2000) 191}}, \href{https://arxiv.org/abs/astro-ph/0006305}{[arXiv:astro-ph/0006305]}.
		

\bibitem{annu2}
James M. Lattimer, Madappa  Prakash, ``The Physics of Neutron Stars'', {\hypersetup{urlcolor=vividviolet}\href{https://www.science.org/doi/10.1126/science.1090720}{Science \textbf{304} (2004) 536}}, \href{https://arxiv.org/abs/astro-ph/0405262}{[arXiv:astro-ph/0405262]}.		
		
\bibitem{gen2021a}
Idrus Husin Belfaqih, Harris Maulana, Anto Sulaksono,   ``White Dwarfs and Generalized Uncertainty Principle'', {\hypersetup{urlcolor=vividviolet}\href{https://www.worldscientific.com/doi/abs/10.1142/S0218271821500644}{     Int. J. Mod. Phys. D \textbf{30} (2021)  2150064}}, \href{https://arxiv.org/abs/2104.11774}{[arXiv:gr-qc/2104.11774]}. 			
	
\bibitem{gen2021b}
Adrian G. Abac, Jose Perico H. Esguerra, Roland Emerito S. Otadoy,   ``Modified Structure Equations and Mass-Radius Relations of White Dwarfs Arising From the Linear Generalized Uncertainty Principle'', {\hypersetup{urlcolor=vividviolet}\href{https://www.worldscientific.com/doi/abs/10.1142/S021827182150005X}{Int. J. Mod. Phys. D \textbf{30} (2021)  2150005}}. 		

\bibitem{gen2021c}
Adrian G. Abac, Jose Perico H. Esguerra,   ``Implications of the Generalized Uncertainty Principle on the Walecka Model Equation of State and Neutron Star Structure'', {\hypersetup{urlcolor=vividviolet}\href{https://www.worldscientific.com/doi/abs/10.1142/S0218271821500553}{Int. J. Mod. Phys. D \textbf{30} (2021)  2150055}}. 	



\bibitem{gen2021d}
Stefano Viaggiu,   ``A Proposal for Heisenberg's uncertainty Principle and Stur for Curved Backgrounds: An Application to White Dwarf, Neutron Stars and Black Holes'', {\hypersetup{urlcolor=vividviolet}\href{https://iopscience.iop.org/article/10.1088/1361-6382/abc907}{Class. Quantum Grav. \textbf{38} (2021)  025017}}, \href{https://arxiv.org/abs/2012.10103}{[arXiv:gr-qc/2012.10103]}. 			
	
	

\bibitem{gen2021e}
Xin-Dong Du, Chao-Yun Long,   ``Removing The Divergence of Chandrasekhar Limit Caused by Generalized Uncertainty Principle'', {\hypersetup{urlcolor=vividviolet}\href{https://link.springer.com/article/10.1140/epjc/s10052-022-10723-0}{Eur. Phys. Jour. C \textbf{82} (2022)  748}}, \href{https://arxiv.org/abs/2201.04338}{[arXiv:astro-ph.HE/2201.04338]}. 			
	
	
\bibitem{garf}
David Garfinkle,   ``The Planck Mass and the Chandrasekhar Limit'', {\hypersetup{urlcolor=vividviolet}\href{https://aapt.scitation.org/doi/10.1119/1.3110884}{Am. J. Phys. \textbf{77} (2009)  683}}. 	

\bibitem{gibbons1}
Gary W. Gibbons,   ``The Maximum Tension Principle in General Relativity'', {\hypersetup{urlcolor=vividviolet}\href{https://link.springer.com/article/10.1023\%2FA\%3A1022370717626}{Found. Phys. \textbf{32} (2002)  1891}}, \href{https://arxiv.org/abs/hep-th/0210109}{[arXiv:hep-th/0210109]}. 	

\bibitem{gibbons2}
John D. Barrow, Gary W. Gibbons,   ``Maximum Tension: With and Without a Cosmological Constant'', {\hypersetup{urlcolor=vividviolet}\href{https://academic.oup.com/mnras/article/446/4/3874/2892906}{MNRAS \textbf{446} (2015)  3874}}, \href{https://arxiv.org/abs/1408.1820}{[arXiv:gr-qc/1408.1820]}. 	

\bibitem{shiller1}
Christoph Schiller,   ``General Relativity and Cosmology Derived From Principle of Maximum Power or Force'', {\hypersetup{urlcolor=vividviolet}\href{https://link.springer.com/article/10.1007\%2Fs10773-005-4835-2}{ Int. J. Theor. Phys. \textbf{44} (2005)  1629}}. 	

\bibitem{shiller2}
Christoph Schiller,   ``Simple Derivation of Minimum Length, Minimum Dipole Moment and Lack of Space-Time Continuity'', {\hypersetup{urlcolor=vividviolet}\href{https://link.springer.com/article/10.1007\%2Fs10773-005-9018-7}{Int. J. Theor. Phys. \textbf{45} (2006)  213}}. 	

\bibitem{dyson}
Freeman Dyson, in {\it Interstellar Communication}, ed. A.G. Cameron, (New York: Benjamin, 1963), chap. 12.		

		
\bibitem{amelino1}
Won Sang Chung, Hassan Hassanabadi, ``New Generalized Uncertainty Principle From the Doubly Special Relativity'', {\hypersetup{urlcolor=vividviolet}\href{https://www.sciencedirect.com/science/article/pii/S0370269318306373?via\%3Dihub}{Phys. Lett. B \textbf{785} (2018) 127}}, \href{https://arxiv.org/abs/1807.11552}{[arXiv:gr-qc/1807.11552]}.
		
		
\bibitem{amelino2}
Joao Magueijo, Lee Smolin, ``Lorentz Invariance With an Invariant Energy Scale'', {\hypersetup{urlcolor=vividviolet}\href{https://journals.aps.org/prl/abstract/10.1103/PhysRevLett.88.190403}{Phys. Rev. Lett. \textbf{88} (2002) 190403}}, \href{https://arxiv.org/abs/hep-th/0112090}{[arXiv:hep-th/0112090]}.



\bibitem{amelino3}
Joao Magueijo, Lee Smolin, ``String Theories With Deformed Energy Momentum Relations, and a Possible Non-Tachyonic Bosonic String'', {\hypersetup{urlcolor=vividviolet}\href{https://journals.aps.org/prd/abstract/10.1103/PhysRevD.71.026010}{Phys. Rev. D \textbf{71} (2005) 026010}}, \href{https://arxiv.org/abs/hep-th/0401087}{[arXiv:hep-th/0401087]}.
		
		
\bibitem{amelino4}
Jose Luis Cort\'es, Jorge Gamboa, ``Quantum Uncertainty in Doubly Special Relativity'', {\hypersetup{urlcolor=vividviolet}\href{https://journals.aps.org/prd/abstract/10.1103/PhysRevD.71.065015}{Phys. Rev. D \textbf{71} (2005) 065015}}, \href{https://arxiv.org/abs/hep-th/0405285}{[arXiv:hep-th/0405285]}.
		
	

\bibitem{jeans}
Zhong-Wen Feng, Guansheng He, Xia Zhou, Xueling Mu, Shi-Qi Zhou, ``Higher-Order Generalized Uncertainty Principle Corrections to the Jeans Mass'', {\hypersetup{urlcolor=vividviolet}\href{https://link.springer.com/article/10.1140\%2Fepjc\%2Fs10052-021-09549-z}{Eur. Phys. J. C \textbf{81} (2021) 754}}, \href{https://arxiv.org/abs/2006.01698}{[arXiv:physics.gen-ph/2006.01698]}.
		
		
\bibitem{Bok}
Ryo Kandori, Yasushi Nakajima, Motohide Tamura, Ken'ichi Tatematsu, Yuri Aikawa, Takahiro Naoi, Koji Sugitani, Hidehiko Nakaya, Takahiro Nagayama, Tetsuya Nagata, Mikio Kurita, Daisuke Kato, Chie Nagashima, Shuji Sato, ``Near Infrared Imaging Survey of Bok Globules: Density Structure'', {\hypersetup{urlcolor=vividviolet}\href{https://iopscience.iop.org/article/10.1086/444619}{Astron. J. \textbf{130} (2005) 2166}}, \href{https://arxiv.org/abs/astro-ph/0506205}{[arXiv:astro-ph/0506205]}.		
		
		


 
 
 	
\bibitem{phasespace}
Bilel Hamil, Bekir Can L\"utf\"uo\v{g}lu,   ``New Higher-Order Generalized Uncertainty Principle: Applications'', {\hypersetup{urlcolor=vividviolet}\href{https://link.springer.com/article/10.1007\%2Fs10773-021-04853-6}{ Int. J. Theor. Phys.  \textbf{60} (2021)  2790}}, \href{https://arxiv.org/abs/2009.13838v1}{[arXiv:gr-qc/2009.13838]}.


\bibitem{liouv}
Lay Nam Chang, Djordje Minic, Naotoshi Okamura, Tatsu Takeuchi,   ``Effect of the Minimal Length Uncertainty  Relation on the Density of States and the Cosmological Constant Problem'', {\hypersetup{urlcolor=vividviolet}\href{https://journals.aps.org/prd/abstract/10.1103/PhysRevD.65.125028}{Phys. Rev. D  \textbf{65} (2002)  125028}}, \href{https://arxiv.org/abs/hep-th/0201017}{[arXiv:hep-th/0201017]}.

\bibitem{rgupc}
Marcos A. Anacleto, Francisco A. Brito, J.A.V. Campos, Eduardo Passos,   ``Quantum-Corrected Scattering and Absorption of a Schwarzschild Black Hole With GUP'', {\hypersetup{urlcolor=vividviolet}\href{https://www.sciencedirect.com/science/article/pii/S037026932030633X?via\%3Dihub}{Phys. Lett. B  \textbf{810} (2020)  135830}}, \href{https://arxiv.org/abs/2003.13464}{[arXiv:gr-qc/2003.13464]}.
		

\bibitem{halo}
Alis J. Deason, Azadeh Fattahi, Carlos S. Frenk, Robert J. J. Grand,  Kyle A. Oman, Shea Garrison-Kimmel, Christine M. Simpson, Julio F. Navarro,   ``The Edge of the Galaxy'', {\hypersetup{urlcolor=vividviolet}\href{https://academic.oup.com/mnras/article-abstract/496/3/3929/5858908?redirectedFrom=fulltext}{ 	MNRAS  \textbf{496} (2020)  3929}}, \href{https://arxiv.org/abs/2002.09497}{[arXiv:astro-ph.GA/2002.09497]}.

\bibitem{iocco}
Jakob Henrichs, Margherita Lembo, Fabio Iocco, Luca Amendola,   ``Testing Gravity With the Milky Way: Yukawa Potential'', {\hypersetup{urlcolor=vividviolet}\href{https://journals.aps.org/prd/abstract/10.1103/PhysRevD.104.043009}{Phys. Rev. D  \textbf{104} (2021)  043009}}, \href{https://arxiv.org/abs/2010.15190}{[arXiv:astro-ph.GA/2010.15190]}.



\bibitem{nuclear}
Ilham Prasetyo, I. H. Belfaqih, A. B. Wahidin, Agus  Suroso, Anto Sulaksono,   ``Minimal length, nuclear matter, and neutron stars'', {\hypersetup{urlcolor=vividviolet}\href{https://link.springer.com/article/10.1140/epjc/s10052-022-10849-1}{ EPJC  \textbf{82} (2022)  884}}, \href{https://arxiv.org/abs/2210.11727}{[arXiv:nucl-th/2210.11727]}.

\bibitem{NL1}
Pisin Chen, Yen Chin Ong, Dong-han Yeom,   ``Generalized Uncertainty Principle: Implications for Black Hole Complementarity'', {\hypersetup{urlcolor=vividviolet}\href{https://link.springer.com/article/10.1007\%2FJHEP12\%282014\%29021}{ JHEP  \textbf{12} (2014)  021}}, \href{https://arxiv.org/abs/1408.3763v2}{[arXiv:hep-th/1408.3763]}.
		
		
\bibitem{NL2}
Ramy Brustein, Georgi Dvali, Gabriele Veneziano,   ``A Bound on the Effective Gravitational Coupling From Semiclassical Black Holes'', {\hypersetup{urlcolor=vividviolet}\href{https://iopscience.iop.org/article/10.1088/1126-6708/2009/10/085}{JHEP  \textbf{0910} (2009)  085}}, \href{https://arxiv.org/abs/0907.5516}{[arXiv:hep-th/0907.5516]}.
		
		
\bibitem{NL3}
Rui Vilela Mendes,   ``Space-Time: Commutative or Noncommutative? Two Length Scales of Noncommutativity'', {\hypersetup{urlcolor=vividviolet}\href{https://journals.aps.org/prd/abstract/10.1103/PhysRevD.99.123006}{ Phys. Rev. D  \textbf{99} (2019)   123006}}, \href{https://arxiv.org/abs/1901.01613v2}{[arXiv:hep-th/1901.01613]}.



\bibitem{bag1}
Mark Alford, Matt Braby, Mark Paris, Sanjay Reddy, ``Hybrid Stars that Masquerade as Neutron Stars'', {\hypersetup{urlcolor=vividviolet}\href{https://iopscience.iop.org/article/10.1086/430902}{Astrophys. J.  \textbf{629} (2005) 969}}, \href{https://arxiv.org/abs/nucl-th/0411016}{[arXiv:nucl-th/0411016]}.

\bibitem{bag2}
Andrew W. Steiner, James M. Lattimer, Edward F. Brown,  ``The Neutron Star Mass-Radius Relation and the Equation of State of Dense Matter'', {\hypersetup{urlcolor=vividviolet}\href{https://iopscience.iop.org/article/10.1088/2041-8205/765/1/L5}{Astrophys. J. Lett.  \textbf{765} (2013) L5}}, \href{https://arxiv.org/abs/1205.6871}{[arXiv:nucl-th/1205.6871]}.

\bibitem{chatelier}
Norman K. Glendenning, ``Lower Limit on Radius as a Function of Mass for Neutron Stars'', {\hypersetup{urlcolor=vividviolet}\href{https://journals.aps.org/prl/abstract/10.1103/PhysRevLett.85.1150}{Phys. Rev. Lett.  \textbf{85} (2000) 1150}}, \href{https://arxiv.org/abs/astro-ph/0003244}{[arXiv:astro-ph/0003244]}.


\bibitem{wein}
Steven Weinberg,   {\it Gravitation and Cosmology: Principles and Applications of the General Theory of Relativity}, (John Wiley \& Sons 1972).

\bibitem{buch}
Hans Adolf Buchdahl,   ``General Relativistic Fluid Spheres'', {\hypersetup{urlcolor=vividviolet}\href{https://journals.aps.org/pr/abstract/10.1103/PhysRev.116.1027}{Phys. Rev.  \textbf{116} (1959)  1027}}.

\bibitem{ref14}
Tsutomu Kobayashi, Yuki Watanabe, Daisuke Yamauchi,   ``Breaking of Vainshtein screening in Scalar-Tensor Theories beyond Horndeski'',  {\hypersetup{urlcolor=vividviolet}\href{https://journals.aps.org/prd/abstract/10.1103/PhysRevD.91.064013}{Phys. Rev. D  \textbf{91} (2015)  064013}}, \href{https://arxiv.org/abs/1411.4130}{[arXiv:gr-qc/1411.4130]}.


\bibitem{ref15}
Luca Amendola, Martin Kunz, Ippocratis D. Saltas, Ignacy Sawicki, ``The Fate of large-scale Structure in Modified Gravity
after GW170817 and GRB170817A'', {\hypersetup{urlcolor=vividviolet}\href{https://journals.aps.org/prl/abstract/10.1103/PhysRevLett.120.131101}{Phys. Rev. Lett.  \textbf{120} (2018) 131101}}, \href{https://arxiv.org/abs/1711.04825}{[arXiv:astro-ph.CO/1711.04825]}.

\bibitem{jcapref}
Ippocratis D. Saltas, Ignacy Sawicki, Ilidio Lopes, ``White Dwarfs and Revelations'', {\hypersetup{urlcolor=vividviolet}\href{https://iopscience.iop.org/article/10.1088/1475-7516/2018/05/028}{JCAP  \textbf{05} (2018) 028}}, \href{https://arxiv.org/abs/1803.00541}{[arXiv:astro-ph.CO/1803.00541]}.

\bibitem{ref18}
Marco Crisostomi, Kazuya Koyama, ``Vainshtein mechanism after GW170817'', {\hypersetup{urlcolor=vividviolet}\href{https://journals.aps.org/prd/abstract/10.1103/PhysRevD.97.021301}{Phys. Rev. D  \textbf{97} (2018) 021301(R)}}, \href{https://arxiv.org/abs/1711.06661}{[arXiv:astro-ph.CO/1711.06661]}.

\bibitem{ref19}
Alexandru Dima, Filippo Vernizzi, ``Vainshtein Screening in Scalar-Tensor Theories before and after GW170817: Constraints on Theories beyond Horndeski'', {\hypersetup{urlcolor=vividviolet}\href{https://journals.aps.org/prd/abstract/10.1103/PhysRevD.97.101302}{Phys. Rev. D  \textbf{97} (2018)  101302(R)}}, \href{https://arxiv.org/abs/1712.04731}{[arXiv:gr-qc/1712.04731]}.

\bibitem{valoriY}
Rajeev Kumar Jain, Chris Kouvaris, Niklas Gr\o nlund Nielsen, ``White Dwarf Critical Tests for Modified Gravity'', {\hypersetup{urlcolor=vividviolet}\href{https://journals.aps.org/prl/abstract/10.1103/PhysRevLett.116.151103}{Phys. Rev. Lett.  \textbf{116} (2016)  151103}}, \href{https://arxiv.org/abs/1512.05946}{[arXiv:astro-ph.CO/1512.05946]}.

\bibitem{integrals1}
Sheise M. de Carvalho, Michael Rotondo, Jorge A. Rueda, Remo Ruffini, ``The relativistic Feynman-Metropolis-Teller Treatment at finite Temperatures'', {\hypersetup{urlcolor=vividviolet}\href{https://journals.aps.org/prc/abstract/10.1103/PhysRevC.89.015801}{Phys. Rev. C  \textbf{89} (2014)  015801}}, \href{https://arxiv.org/abs/1312.2434}{[arXiv:astro-ph.SR/1312.2434]}.



\bibitem{gabriele}
Gabriele Barca, Eleonora Giovannetti, Giovanni Montani,   ``Comparison of the Semiclassical and Quantum Dynamics of the Bianchi I Cosmology in the Polymer and GUP Extended Paradigms'',  {\hypersetup{urlcolor=vividviolet}\href{https://www.worldscientific.com/doi/10.1142/S0219887822500979}{Int. Jour. Geom. Meth. Mod. Phys.  \textbf{19} (2022)  2250097}}, \href{https://arxiv.org/abs/2112.08905}{[arXiv:gr-qc/2112.0890]}.

\bibitem{dark1}
Rahul Ghosh, Surajit Chattopadhyay, Ujjal Debnath,   ``A Dark Energy Model with Generalized Uncertainty Principle in the Emergent, Intermediate and Logamediate Scenarios of the Universe'', {\hypersetup{urlcolor=vividviolet}\href{https://link.springer.com/article/10.1007\%2Fs10773-011-0939-z}{  	Int. J. Theor. Phys.  \textbf{51} (2012)  589}}, \href{https://arxiv.org/abs/1105.4538}{[arXiv:gr-qc/1105.4538]}.
		
		
\bibitem{dark2}
Mahdi Rashki, Majid Fathi, Behrang Mostaghel, Sahram Jalalzadeh,   ``Interacting Dark Side of Universe Through Generalized Uncertainty Principle'', {\hypersetup{urlcolor=vividviolet}\href{https://www.worldscientific.com/doi/abs/10.1142/S0218271819500810}{ Int. J. Mod. Phys. D  \textbf{28} (2019)  1950081}}, \href{https://arxiv.org/abs/1901.05766}{[arXiv:gr-qc/1901.05766]}.
		
		
\bibitem{dark3}
Yong-Wan Kim, Hyung Won Lee, Yun Soo Myung, Mu-In Park,   ``New Agegraphic Dark Energy Model With Generalized Uncertainty Principle'', {\hypersetup{urlcolor=vividviolet}\href{https://www.worldscientific.com/doi/abs/10.1142/S021773230802848X}{Mod. Phys. Lett. A  \textbf{23} (2008)  3049}}, \href{https://arxiv.org/abs/0803.0574}{[arXiv:gr-qc/0803.0574]}.
		
		
		
\bibitem{dark4}
Shahram Jalalzadeh, Mohammad Ali Gorji, Kourosh Nozari,   ``Deviation from the Standard Uncertainty Principle and the Dark Energy Problem'', {\hypersetup{urlcolor=vividviolet}\href{https://link.springer.com/article/10.1007\%2Fs10714-013-1632-8}{ Gen. Rel. Grav.  \textbf{46} (2014)  1632}}, \href{https://arxiv.org/abs/1310.8065}{[arXiv:gr-qc/1310.8065]}.


	
\bibitem{iorio}
Alfredo Iorio, Gaetano Lambiase, Pablo Pais, Fabio Scardigli,   ``Generalized Uncertainty Principle in Three-Dimensional Gravity and the BTZ Black Hole'', {\hypersetup{urlcolor=vividviolet}\href{https://journals.aps.org/prd/abstract/10.1103/PhysRevD.101.105002}{ Phys. Rev. D  \textbf{101} (2020)  105002}}, \href{https://arxiv.org/abs/1910.09019}{[arXiv:hep-th/1910.09019]}.
	

	
\bibitem{scardigli}
Fabio Scardigli,   ``Generalized Uncertainty Principle in Quantum Gravity from Micro-Black Hole Gedanken Experiment'', {\hypersetup{urlcolor=vividviolet}\href{https://www.sciencedirect.com/science/article/abs/pii/S0370269399001677?via\%3Dihub}{ Phys. Lett. B \textbf{452} (1999)  39}}, \href{https://arxiv.org/abs/hep-th/9904025}{[arXiv:hep-th/9904025]}.
	









\end{thebibliography}
\end{document}